\documentclass[12pt]{article}
\usepackage{a4wide,cite,rotate,epsf}

\newcommand{\cK}{{\cal K}}
\newcommand{\cL}{{\cal L}}
\newcommand{\cM}{{\cal M}}
\newcommand{\cO}{{\cal O}}

\newcommand{\cR}{{\cal R}}

\newcommand{\nn}{\nonumber}

\newcommand{\IM}{\mbox{\rm Im}}
\newcommand{\RE}{\mbox{\rm Re}}
\newcommand{\eqn}[1]{(\ref{#1})}

\newcommand{\mev}{\mbox{\rm MeV}}
\newcommand{\gev}{\mbox{\rm GeV}}

\newcommand{\DelKP}{\Delta_{K\pi}}
\newcommand{\SigKP}{\Sigma_{K\pi}}

\newcommand{\msmall}[1]{\mbox{$#1$}}
\newcommand{\tvs}{\vbox{\vskip 6mm}}
\newcommand{\smvs}{\vbox{\vskip 8mm}}

\newcommand{\newsection}[1]{\section{#1}\setcounter{equation}{0}}
\newcommand{\dotM}[1]{{\stackrel{\phantom{2}\,\;\;0\;\,2\phantom{#1}}{M_{#1}}}}

\begin{document}


\begin{titlepage}
\phantom{XXX}
\vspace{-1cm}
\begin{flushright}
{\small\sf IFIC/00-28} \\[-1mm]
{\small\sf FTUV/00-0605} \\[-1mm]
{\small\sf HD-THEP-00-7} \\[-1mm]
{\small\sf FZJ-IKP(TH)-00-14} \\[1cm]
\end{flushright}

\begin{center}
{\LARGE\bf\boldmath S-wave $K\pi$ scattering in chiral \\[3mm]
 perturbation theory with resonances
 \unboldmath}\\[1cm]

{\normalsize\bf Matthias Jamin${}^{1}$, Jos\'e Antonio Oller${}^{2}$ and
 Antonio Pich${}^{3}$} \\[4mm]

{\small\sl ${}^{1}$ Institut f\"ur Theoretische Physik, Universit\"at
           Heidelberg,} \\
{\small\sl Philosophenweg 16, D-69120 Heidelberg, Germany}\\
{\small\sl ${}^{2}$ Forschungszentrum J\"ulich, Institut f\"ur
           Kernphysik (Theorie),} \\
{\small\sl D-52425 J\"ulich, Germany}\\
{\small\sl ${}^{3}$ Departament de F\'{\i}sica Te\`orica - IFIC,
           Universitat de Val\`encia - CSIC,}\\
{\small\sl Apt. Correus 2085, E-46071 Val\`encia, Spain} \\[20mm]
\end{center}

{\bf Abstract:}
We present a detailed analysis of S-wave $K\pi$ scattering up to $2\;\gev$,
making use of the resonance chiral Lagrangian predictions together with a
suitable unitarisation method. Our approach incorporates known theoretical
constraints at low and high energies. The present experimental status, with
partly conflicting data from different experiments, is discussed. Our analysis
allows to resolve some experimental ambiguities, but better data are needed
in order to determine the cross-section in the higher-energy range. Our best
fits are used to determine the masses and widths of the relevant scalar
resonances in this energy region.

\vfill


\\
\phantom{Keywords:} Inelastic scattering: two-particle final states \\[6mm]

\end{titlepage}

\newpage
\setcounter{page}{1}


\newsection{Introduction}

Chiral Perturbation Theory ($\chi$PT) \cite{wei:79,gl:84,gl:85,mei:93,eck:95,
pic:95,bkm:95} provides a very powerful framework to study the low-energy
dynamics of the lightest pseudoscalar octet. The QCD chiral symmetry
constraints imply the pseudo-Goldstone nature of the $\pi$, $K$ and $\eta$
mesons and determine their interactions at very low energies.

To describe the resonance region around 1~GeV, additional non-trivial dynamical
information is needed. One can construct a chiral-symmetric Effective Field
Theory with resonance fields as explicit degrees of freedom \cite{egpr:89,
eglpr:89}. Although not as predictive as the standard chiral Lagrangian for
the pseudo-Goldstone mesons, the resonance chiral Lagrangian \cite{egpr:89}
turns out to provide interesting results, once some short-distance dynamical
QCD constraints are taken into account \cite{eglpr:89}. At tree level, this
Lagrangian encodes the large-$N_c$ properties of QCD \cite{tho:74,wit:79},
in the approximation of keeping only the dominant lowest-mass resonance
multiplets \cite{ppr:98}.

The $\chi$PT loops incorporate the unitarity field theory constraints in a
perturbative way, order by order in the momentum expansion. In the resonance
region, the low-energy expansion breaks down and some kind of resummation
of those chiral loops is required to satisfy unitarity \cite{tru:88,dht:90,
ksw:95,oo:97,gp:97,gue:98,oop:99,oo:99,oor:00}. This effect appears to be
crucial for a correct understanding of the scalar sector, because the S-wave
rescattering of two pseudoscalars is very strong.

In this paper, we present an attempt to describe S-wave $K\pi$ scattering up
to 2~GeV, making use of the resonance chiral Lagrangian predictions together
with a suitable unitarisation method. The analysis is far from trivial, owing
to the very unsatisfactory experimental status of the scalar sector, with
partly conflicting data from different experiments. Even the masses and widths
of the dominant scalar resonances are not clearly established, since the large
unitarity corrections tend to mask the resonance behaviour.
  
A good understanding of the scalar isospin-$1/2$ system is required in order
to improve the determination of the strange quark mass from QCD sum rules for
the strangeness-changing scalar current \cite{nprt:83,mm:93,cdps:95,cfnp:97,
jam:97,bgm:98}, since the phenomenological parametrisation of the corresponding
spectral function can be related to the $K\pi$ scalar form factor. We plan to
discuss this form factor in detail in a subsequent publication \cite{jop:00b}.

In section 2 we collect the one-loop $\chi$PT results for the $K\pi$
scattering amplitude, including scalar and vector resonance contributions.
Our unitarisation approach for the elastic channel is discussed in section 3.
Section 4 analyses the experimental data up to 1.3 GeV and several
$\chi$PT-inspired fits are performed. At higher energies, one needs to include
inelastic channels; the dominant two-body $K\pi\to K\eta, K\eta'$ modes are
incorporated in section 5, and used in section 6 to fit the higher-energy
data up to 2 GeV. Our fits determine the resonance poles of the T matrix in
the complex plane; this is worked out in section 7, where we compare results
for the resonance masses and widths from different fits. Our conclusions are
briefly summarised in section 8.  To simplify the presentation, we refer to
refs. \cite{mei:93,eck:95,pic:95,bkm:95,egpr:89,eglpr:89} for details on the
chiral formalism and notations, while some technical aspects and cumbersome
formulae are relegated into appendices.

\newsection{\boldmath Chiral expansion for the $K\pi$ scattering amplitudes
 \unboldmath}

Up to one loop in the chiral expansion the $K\pi$ scattering amplitudes
have been calculated by Bernard, Kaiser and Mei\ss ner \cite{bkm:91a} and
resonance contributions have been included by the same authors in
\cite{bkm:91b}. We shall just review the main expressions for the
$K\pi$ amplitudes to make our work selfcontained.

Since the pion and the kaon have isospin $1$ and $1/2$ respectively,
there exist two independent $K\pi$ amplitudes $T_{K\pi}^{1/2}$ and
$T_{K\pi}^{3/2}$ (in the following, for simplicity, we shall drop the
subscript ``$K\pi$''). The charge-two process $K^+\pi^+\rightarrow K^+\pi^+$
is purely $I=3/2$ whereas the process $K^+\pi^-\rightarrow K^+\pi^-$
contains both $I=1/2$ and $I=3/2$ components. Both amplitudes depend on
the Mandelstam variables $s$, $t$ and $u$ and are related by
$s\leftrightarrow u$ crossing. From this relation one finds\footnote{For
further details the reader is referred to refs. \cite{bkm:91a, lan:78}.}
\begin{equation}
\label{eq:2.1}
T^{1/2}(s,t,u) \;=\; \frac{3}{2}\,T^{3/2}(u,t,s)-\frac{1}{2}\,T^{3/2}(s,t,u)\,.
\end{equation}
Therefore it is sufficient to give here only the $I=3/2$ amplitude $T^{3/2}$.

At order $p^4$ in the $\chi$PT expansion including resonances, $T^{3/2}$
can be decomposed as follows
\begin{equation}
\label{eq:2.2}
T^{3/2} \;=\; T^{(2)} + T^{(4)}_R + T^{(4)}_U + T^{(4)}_T \,,
\end{equation}
where $T^{(2)}$ is the leading order $p^2$ contribution already known
from current algebra \cite{wei:66}. $T^{(4)}_R$, $T^{(4)}_U$ and $T^{(4)}_T$
are order $p^4$ resonance, unitarity and tadpole corrections respectively.
We have assumed that at the resonance-mass scale the low-energy constants
$L_i$ in the order $p^4$ chiral Lagrangian are saturated by resonance
exchange \cite{egpr:89,eglpr:89}. The tree-level contribution $T^{(2)}$ is
given by
\begin{equation}
\label{eq:2.3}
T^{(2)}(s) \;=\; \frac{1}{2f_K f_\pi}\,\big(\,M_K^2 + M_\pi^2 - s\,\big) \,.
\end{equation}
In writing eq.~\eqn{eq:2.3}, we have absorbed some order $p^4$ corrections
into $T^{(2)}$ by replacing $1/f_\pi^2$ with $1/(f_K f_\pi)$.

Below, we list explicit expressions for the order $p^4$ contributions
$T^{(4)}_R$, $T^{(4)}_U$ and $T^{(4)}_T$ to the amplitude $T^{3/2}$ in
$\chi$PT with resonances. The resonance contribution $T^{(4)}_R(s,t,u)$
is given by
\begin{eqnarray}
\label{eq:2.4}
T^{(4)}_R &=& \frac{1}{f_K^2 f_\pi^2} \Biggl\{\, \frac{G_V^2}{2}
\Biggl[ \frac{t(u-s)}{M_\rho^2-t} + \frac{u(t-s)+\DelKP^2}{M_{K^*}^2-u}
\Biggr] \nn \\
\tvs
&& +\,\frac{1}{M_{K^*_0}^2 -u} \Big[c_d\,u - (c_d-c_m) \SigKP\Big]^2
\, - \frac{c_m^2}{M_S^2} \,\DelKP^2 \nn \\
\tvs
&& +\,\frac{1}{3}\Biggl[\frac{4}{M_{S_1}^2-t} - \frac{1}{M_S^2-t}
\Biggr] \Big[c_d\,t - 2(c_d-c_m) M_K^2\Big]\Big[c_d\,t - 2(c_d-c_m) M_\pi^2
\Big] \Biggr\} \,,
\end{eqnarray}
where we defined the constants $\DelKP\equiv(M_K^2-M_\pi^2)$ and
$\SigKP\equiv(M_K^2+M_\pi^2)$. $G_V$ as well as $c_d$ and $c_m$ are
coupling constants of the vector and scalar mesons in the chiral
resonance Lagrangian of refs. \cite{egpr:89,eglpr:89} respectively.
For the coupling constants of the scalar singlet meson $S_1$,
${\tilde c}_d$ and ${\tilde c}_m$, we have already used the large-$N_c$
constraints $\tilde c_d=c_d/\sqrt{3}$ and $\tilde c_m=c_m/\sqrt{3}$.
$M_S$ refers to the mass of a generic scalar octet meson and we shall
always employ the large-$N_c$ result $M_{S_1}=M_S$.

The unitarity correction $T^{(4)}_U(s,t,u)$ has originally been calculated
in ref. \cite{bkm:91a} and takes the form
\begin{eqnarray}
\label{eq:2.5}
T^{(4)}_U &=& \frac{1}{4f_K^2 f_\pi^2}\biggl\{\,t(u-s)\Big[2M_{\pi\pi}^r(t)+
M_{KK}^r(t)\Big]+\msmall{\frac{3}{2}}\Big\{(s-t)\Big[L_{\pi K}(u)+L_{K\eta_8}(u)
\nn \\
\tvs
&& -\,u\Big(M_{\pi K}^r(u)+M_{K\eta_8}^r(u)\Big)\Big]+
\DelKP^2\,\Big(M_{\pi K}^r(u)+M_{K\eta_8}^r(u)\Big)\Big\} \nn \\
\tvs
&& +\,\msmall{\frac{1}{2}}\DelKP\Big[(5u-2\SigKP)K_{\pi K}(u)+
(3u-2\SigKP)K_{K\eta_8}(u)\Big]+(s-\SigKP)^2 J_{\pi K}^r(s) \nn \\
\tvs
&& +\,\msmall{\frac{1}{8}}\Big[11u^2-12\SigKP\,u+4\SigKP^2\Big] J_{\pi K}^r(u)
+\msmall{\frac{3}{8}}\Big(u-\msmall{\frac{2}{3}}\SigKP\Big)^2 J_{K\eta_8}^r(u)
\nn \\
\tvs
&& +\,\msmall{\frac{1}{2}}\,t(2t-M_\pi^2) J_{\pi\pi}^r(t)
+\,\msmall{\frac{3}{4}}\,t^2 J_{KK}^r(t) + \msmall{\frac{1}{2}}M_\pi^2
\Big(t-\msmall{\frac{8}{9}}M_K^2 \Big) J_{\eta_8\eta_8}^r(t)\,\biggr\} \,.
\end{eqnarray}
We have already corrected for two misprints which appeared in the
original publication \cite{bkm:91a} and the explicit form of the loop
functions can be found in appendix~A. The unitarity correction has also
recently been checked independently in ref. \cite{roe:99}.

Finally, the tadpole contribution $T^{(4)}_T$ reads
\begin{equation}
\label{eq:2.6}
T^{(4)}_T \;=\; -\,\frac{\DelKP}{16f_K f_\pi}\,\Big[\,3\mu_\pi - 2\mu_K
-\mu_{\eta_8}\,\Big] \,,
\end{equation}
with $\mu_P$ being functions which depend on the renormalisation scale
$\mu$ and are given by
\begin{equation}
\label{eq:2.7}
\mu_P \;=\; \frac{M_P^2}{32\pi^2 f_\pi^2}\,\ln\Biggl(\frac{M_P^2}{\mu^2}
\Biggr) \,.
\end{equation}
It has been assumed that the resonance contribution saturates the
low-energy constants $L_i$ and thus, in accordance with ref. \cite{egpr:89},
in the following we always set the renormalisation scale $\mu=M_\rho$,
where this saturation is found to take place. Both, the resonance as well
as the tadpole contribution above, take different forms compared to the
original papers \cite{bkm:91a,bkm:91b}, because we chose to write the
tree-level term \eqn{eq:2.3} differently. However, we have independently
calculated these terms finding agreement with \cite{bkm:91b}.

Since in this work we are only interested in the S-wave $K\pi$
scattering amplitude, the partial wave projection still has to be
performed. Generally, the partial wave amplitudes $t_l^I(s)$ are
given by
\begin{equation}
\label{eq:2.8}
t_l^I(s) \;=\; \frac{1}{32\pi}\int\limits_{-1}^{1} dz \,
P_l(z) \, T_{K\pi}^I(s,z) \,,
\end{equation}
where $l$ is the total angular momentum, $z=\cos\theta$ is the scattering
angle in the c.m. system and $P_l(z)$ are Legendre polynomials \cite{gr:80}.

\newsection{Unitarisation of the amplitudes}

A numerical inspection of eq.~\eqn{eq:2.2} shows that already at energies of
the order of $1\;\gev$ the unitarity correction $T^{(4)}_U$ is rather large.
Since we aim at describing S-wave $K\pi$ scattering well above $1\;\gev$, we
have to resort to methods which unitarise the partial wave amplitudes. The
approach we are going to use is based on the N/D method \cite{cm:60} and has
been developed in refs. \cite{oo:99,oll:00,oor:00}. It can be derived by
working out the partial wave amplitudes from a perturbative loop expansion
in terms of the unphysical cuts. The generated partial waves are then matched
with the ones derived from next-to-leading order $\chi$PT \cite{gl:85}
including in addition the explicit exchange of resonances \cite{egpr:89}. 

Restricting oneself to two-body unitarity and neglecting the unphysical cut
contributions, with the help of the N/D method \cite{cm:60} in ref.
\cite{oo:99} the corresponding general structure of a partial wave amplitude
was derived. This structure was shown to match with the local tree-level
$\chi$PT amplitudes plus the $s$-channel exchange of explicit resonance fields.
It was also estimated that the contributions of the unphysical cuts in the
physical region below $1\;\gev$ for the $I=1/2$ meson-meson S-waves amount
just to a few percent. Then, in ref. \cite{oll:00,oor:00} the generalisation
of this approach to include the unphysical cuts up to one loop
in $\chi$PT was established. Note that the present expansion, treating the
unphysical cut contributions as small quantities, can be implemented as a
chiral loop expansion of these contributions. This means that at the tree
level our amplitudes are crossing symmetric, e.g. resonance exchanges are
included in all the $s$, $t$ and $u$-channels, and that the resulting partial
waves amplitudes fulfil the N/D equations exactly in the right hand cut and
up to one loop in $\chi$PT for the unphysical cuts. This is precisely the
order required to match with the chiral expressions of ref. \cite{bkm:91b},
where the local and pole terms \cite{gl:85,egpr:89} were supplied with loops
calculated at ${\cal O }(p^4)$ in $\chi$PT. In refs. \cite{oll:00,mo:00} the
method described above has already been used to study the strongly interacting
Higgs sector and elastic $\pi N$ scattering, respectively. 

Up to the considered order, we can write the unitarised partial wave
amplitude $\tilde{t}^I_l(s)$ as \cite{oor:00}:\footnote{For more details,
we refer the reader to section 3.2 of the review \cite{oor:00}.}
\begin{equation}
\label{eq:3.1}
\tilde t_l^I(s) \;=\; \frac{N_l^I(s)}{\Big(1-g_{K\pi}^I(s)\,N_l^I(s)\Big)} \,,
\end{equation}
with 
\begin{equation}
\label{eq:3.2}
N_l^I(s) \;=\; t_l^I(s) - g_{K\pi}^I(s) \Big(t_l^{I\,(2)}(s)\Big)^2 \,,
\end{equation}
and
\begin{equation}
\label{eq:3.3}
g^I_{K\pi}(s)=16\pi \bar{J}_{K\pi}(s)+c^I_{K\pi} \,.
\end{equation}
On the right-hand side of eq.~\eqn{eq:3.2}, $t_l^{I\,(2)}(s)$ refers to the
leading ${\cal O }(p^2)$ expression of $t_l^I(s)$, with $t_l^I(s)$ itself being
calculated at the one-loop level in $\chi$PT plus resonances. $\bar{J}(s)$ is
the standard two-particle one-loop integral as given in appendix~A, and the
$c^I_{K\pi}$ are arbitrary constants, not fixed by the unitarisation procedure,
which have to be real in order to satisfy unitarity. Reexpanding \eqn{eq:3.1}
up to one loop, also taking into account eq.~\eqn{eq:3.2}, one can immediately
match the resulting expression to the chiral amplitudes collected in section~2.

It is easy to see that eq.~\eqn{eq:3.1} satisfies elastic unitarity above
the $K\pi$ threshold and below the inelastic ones, namely:
\begin{equation}
\label{eq:3.4}
\IM\,\tilde t_l^I(s) \;=\; \sigma_{K\pi}(s)\,|\tilde t_l^I(s)|^2 \,,
\end{equation}
since by construction $\IM N_l^I(s) \,=\, 0$ in the $K\pi$ physical region and 
\begin{equation}
\label{eq:3.5}
\IM\,g_{K\pi}^I(s+i\,0)\;=\;\sigma_{K\pi}(s) \;\equiv\;
\sqrt{(1-s_+/s)(1-s_-/s)}\,,
\end{equation}
for $s\geq s_+$ where $s_+=(M_K+M_\pi)^2$ and $s_-=(M_K-M_\pi)^2$.

As long as we only consider a single channel, $\tilde{t}_l^I(s)$ and $N^I_l(s)$
are scalar functions. In ref. \cite{oor:00} the rather straightforward
generalisation of eq.~\eqn{eq:3.1} to coupled channels where $\tilde{t}_l^I(s)$
and $N^I_l(s)$ become matrices is also given and we shall come back to it
in section~5 below.

\newsection{Fitting the elastic channel}

Experimentally, it has been observed that the $I=1/2$ $K\pi$ scattering
amplitude turns out to be elastic below roughly $1.3\;\gev$ and the same
was found for the $I=3/2$ amplitude up to $1.9\;\gev$. Therefore, in this
energy range we should be able to describe the experimental data on
$K\pi$ scattering with the unitarised chiral expressions presented in
the previous two sections. First we shall comment on the experimental
data which we included in our fits, before we explain the actual
fitting procedure and our results. 

The two most recent experiments on $K\pi$ scattering have been
performed by Estabrooks et al. \cite{est:78} and Aston et al.
\cite{ast:88,awa:86}. Estabrooks et al. measured charged $K\pi$ scattering
for all four charge combinations and were thus able to separate the
$I=1/2$ and $I=3/2$ components in the elastic region below $1.3\;\gev$.
Aston et al. only considered $K^-\pi^+$ scattering. In this case the
S-wave amplitude in terms of phase shifts and inelasticities is given
as follows:
\begin{equation}
\label{eq:4.1}
A_0 \;=\; a_0\,e^{i\phi_0} \;=\; A_0^{1/2} + \frac{1}{2}\,A_0^{3/2} \;=\;
\frac{1}{2i}\,\biggl(\eta_0^{1/2} e^{2i\delta_0^{1/2}}-1\biggr) +
\frac{1}{4i}\,\biggl(\eta_0^{3/2} e^{2i\delta_0^{3/2}}-1\biggr) \,,
\end{equation}
where $\eta_0^I$ and $\delta_0^I$ are the inelasticities and phase shifts for
the $I=1/2$ and $I=3/2$ channel respectively. In writing eq.~\eqn{eq:4.1}
we have adopted the normalisation of refs.~\cite{est:78,ast:88,awa:86}
for $A_0$. In ref.~\cite{est:78}, $\delta_0^{1/2}$ has been extracted for
$\sqrt s=0.73$ -- $1.30\;\gev$, and $\delta_0^{3/2}$ for $0.73$ -- $1.72\;\gev$.
In this work it was also demonstrated that below $1.3\;\gev$, the scattering
is purely elastic, that is, $\eta_0^{1/2}=\eta_0^{3/2}=1$, whereas it was
found that $\eta_0^{3/2}=1$ in the full energy range. On the other hand
in ref.~\cite{awa:86} only $a_0$ and $\phi_0$ up to $\sqrt s=2.52\;\gev$
were given.

Because the $I=3/2$ component of the scattering amplitude was only measured
directly by Estabrooks et al. \cite{est:78}, we have decided to also
include older data for $\delta_0^{3/2}$ in our fits \cite{jon:73,lin:73,
bak:70,cho:70}. The last three experiments only presented data for the
$K^-\pi^-$ cross section and one still has to extract $\delta_0^{3/2}$ from
the given cross sections. In addition, rather different energy intervals
have been used by the experimental groups \cite{est:78,jon:73,lin:73,bak:70,
cho:70} for calculating cross sections and phase shifts. To be able to
combine the data consistently, we thus also included an error in the energy.
Nevertheless, it turned out that the combined $\chi^2/d.o.f.$ for our fits
was only of the order of one if we increased the error in all the
experiments \cite{est:78,jon:73,lin:73,bak:70,cho:70} by a factor of two.
The increase of errors for these data is implied in the rest of our work.

In terms of the unitarised chiral expressions, the isospin amplitudes $A_0^I$
are given by
\begin{equation}
\label{eq:4.2}
A_0^I(s) \;=\; \frac{\sigma_{K\pi}(s)N_0^I(s)}{\Big(1-g_{K\pi}^I(s)\,N_0^I(s)
\Big)} \,.
\end{equation}
In the elastic region the phase shift $\delta_0^I$ is then related to the
amplitude by $\tan\delta_0^I=\IM A_0^I/\RE A_0^I$. We have fitted all data for
$\delta_0^{1/2}$, $a_0$ and $\phi_0$ up to $1.31\;\gev$ and for
$\delta_0^{3/2}$ up to $1.9\;\gev$ simultaneously. The parameters for the fit
are $M_{K^*_0}$, $c_d$, $c_m$, $c_{K\pi}^{1/2}$ and $c_{K\pi}^{3/2}$. Lacking
further information, and to reduce the number of free parameters, we have set
$M_{S_1}=M_S=M_{K^*_0}$, and we have used the large-$N_c$ estimate
$G_V=f_\pi/\sqrt{2}=65.3\;\mev$ \cite{eglpr:89}. The latter value is in
good agreement with the lowest-order results coming from the decay width of
the $K^*$ ($\rho$) meson which yield $G_V=65\;(67)\;\mev$ and is also
compatible with the original estimate $G_V=53\;\mev$ \cite{egpr:89} which
however already includes chiral loops. All other parameters have been taken
from the Review of Particle Physics \cite{pdg:98}. For the convenience of
the reader we have compiled the constant input parameters in appendix~D.

From our fits we observe that some of the parameters are strongly
correlated and for example rather different values for the couplings of
the scalar resonances can lead to similar fit quality. Thus it is
necessary to have an idea what the values of $c_d$ and $c_m$ should be.
In ref.~\cite{egpr:89} values for these parameters were estimated by
assuming resonance saturation for the low-energy constants in the chiral
Lagrangian. Taking the same approach with the most recent values for
the low-energy constants \cite{abt:99,abt:00}, and an average mass of the
scalar octet $M_S=1.4\;\gev$, we arrive at
\begin{equation}
\label{eq:4.3}
|c_d| \;=\; 30 \pm 10 \;\mev \,, \qquad
|c_m| \;=\; 43 \pm 14 \;\mev \,,
\end{equation}
with $c_d\,c_m>0$, rather close to the original estimate in ref.
\cite{egpr:89}. For the error we have taken one third which is a typical
error for large-$N_c$ estimates of chiral constants. The estimate
$M_S=1.4\;\gev$ tacitly assumes that the correct $I=1$ scalar in this
region is the $a_0(1450)$, and that the $a_0(980)$ is generated dynamically
\cite{wi:82,jphs:95,ac:96,oo:97,efss:98,oo:99}. Note however, that this
question is controversial in the literature
\cite{mor:74,as:91,tor:95}.\footnote{For further discussion and references on
this point see also the {\em note on scalar mesons} in the Review of Particle
Physics \cite{pdg:98}.}

As a first fit, we fix $c_d$ and $c_m$ to the central values presented in
eq.~\eqn{eq:4.3} above. The remaining fit parameters are then found to be
\begin{equation}
\label{eq:4.4}
M_{K^*_0} \;=\; 1.29\;\gev \,, \qquad
c_{K\pi}^{1/2} \;=\; -\,0.094 \,, \qquad
c_{K\pi}^{3/2} \;=\; 1.703 \,,
\end{equation}
with a $\chi^2 = 223.5/117\,d.o.f. = 1.91$. The fits are performed with the
program Minuit \cite{min:94}. Since the fit parameters are highly correlated,
we do not give statistical errors for the fit because they would underestimate
the true uncertainties. We shall return to a discussion of the uncertainties
below.

Next, the $K\pi$ scattering data are fit with the constraint $c_m=c_d$ for
the couplings of the scalar resonances. This constraint is motivated from a
calculation of the scalar form factor including resonances in the large-$N_c$
limit which will be presented in a subsequent publication \cite{jop:00b}.
Requiring the tree-level form factor to vanish at infinity, which should be
a plausible assumption, in the case of one scalar resonance we find that
$c_d = c_m = f_\pi/2$. For the corresponding fit the parameters take the values 
\begin{displaymath}
M_{K^*_0} \;=\; 1.19\;\gev \,, \qquad
c_d \;=\; c_m \;=\; 45.4\;\mev \,,
\end{displaymath}
\begin{equation}
\label{eq:4.5}
c_{K\pi}^{1/2} \;=\; -\,0.376 \,, \qquad
c_{K\pi}^{3/2} \;=\; 1.689 \,.
\end{equation}
The $\chi^2$ for this fit is $210.6/116\,d.o.f. = 1.82$. The resulting values
of $c_d$ and $c_m$ turn out to be surprisingly  close to the large-$N_c$
constraint $f_\pi/2=46.2\;\mev$.

Finally, we perform a fit leaving both $c_d$ and $c_m$ as free parameters.
This fit results in
\begin{displaymath}
M_{K^*_0} \;=\; 1.26\;\gev \,, \qquad
c_d \;=\; 24.8\;\mev \,, \qquad
c_m \;=\; 76.7\;\mev \,,
\end{displaymath}
\begin{equation}
\label{eq:4.6}
c_{K\pi}^{1/2} \;=\; -\,0.191 \,, \qquad
c_{K\pi}^{3/2} \;=\; 1.700 \,,
\end{equation}
and $\chi^2 = 201.1/115\,d.o.f. = 1.75$. Here, $c_m$ turns out to be quite
different from the central estimate given in eq.~\eqn{eq:4.3}. However,
comparing eqs.~\eqn{eq:4.4}, \eqn{eq:4.5} and \eqn{eq:4.6} one sees that
the $\chi^2$ is around two for all three fits. Thus, the difference for
the estimates of $c_d$ and $c_m$ gives an indication about the systematic
uncertainties in their determination.

\begin{figure}[htb]
\centerline{
\rotate[r]{
\epsfysize=15cm
\epsffile{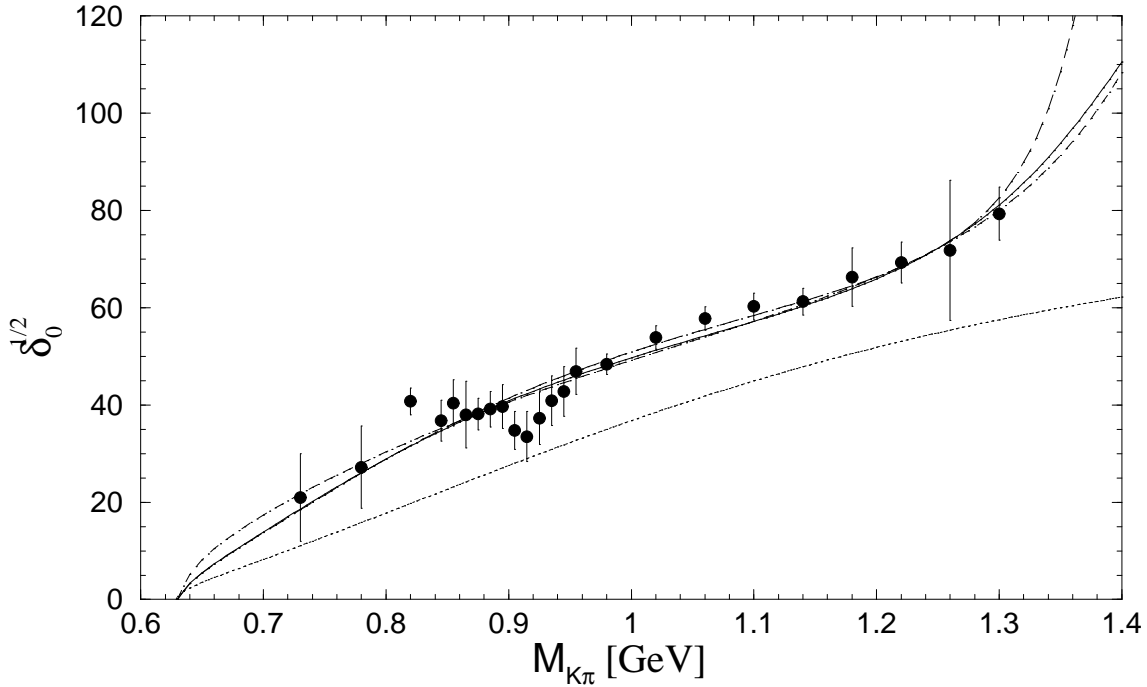} } }
\vspace{-4mm}
\caption[]{Elastic fits for $\delta_0^{1/2}$. Full circles: data of
ref.~\cite{est:78}. Long-dashed line: fit of eq.~\eqn{eq:4.6}; solid line:
fit of eq.~\eqn{eq:4.10}; short-dashed line: fit of eq.~\eqn{eq:4.13};
dotted line: fit of eq.~\eqn{eq:4.14}.
\label{fig:1}}
\end{figure}
\begin{figure}[htb]
\centerline{
\rotate[r]{
\epsfysize=15cm
\epsffile{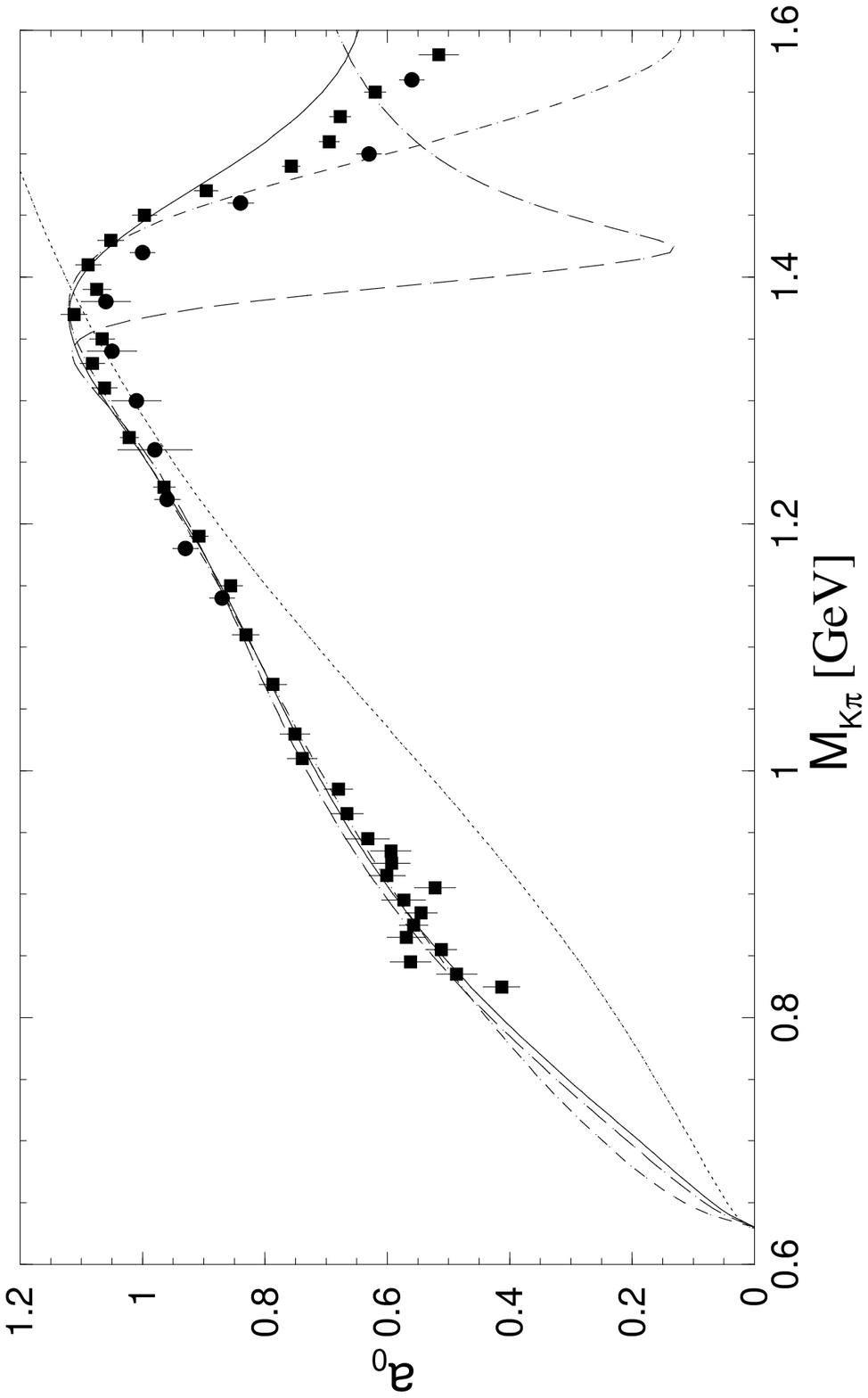} } }
\vspace{-4mm}
\caption[]{Elastic fits for $a_0$. Full circles: data of ref.~\cite{est:78};
full squares: data of ref.~\cite{ast:88,awa:86}. Long-dashed line: fit of
eq.~\eqn{eq:4.6}; solid line: fit of eq.~\eqn{eq:4.10}; short-dashed line:
fit of eq.~\eqn{eq:4.13}; dotted line: fit of eq.~\eqn{eq:4.14}.
\label{fig:2}}
\end{figure}
\begin{figure}[htb]
\centerline{
\rotate[r]{
\epsfysize=15cm
\epsffile{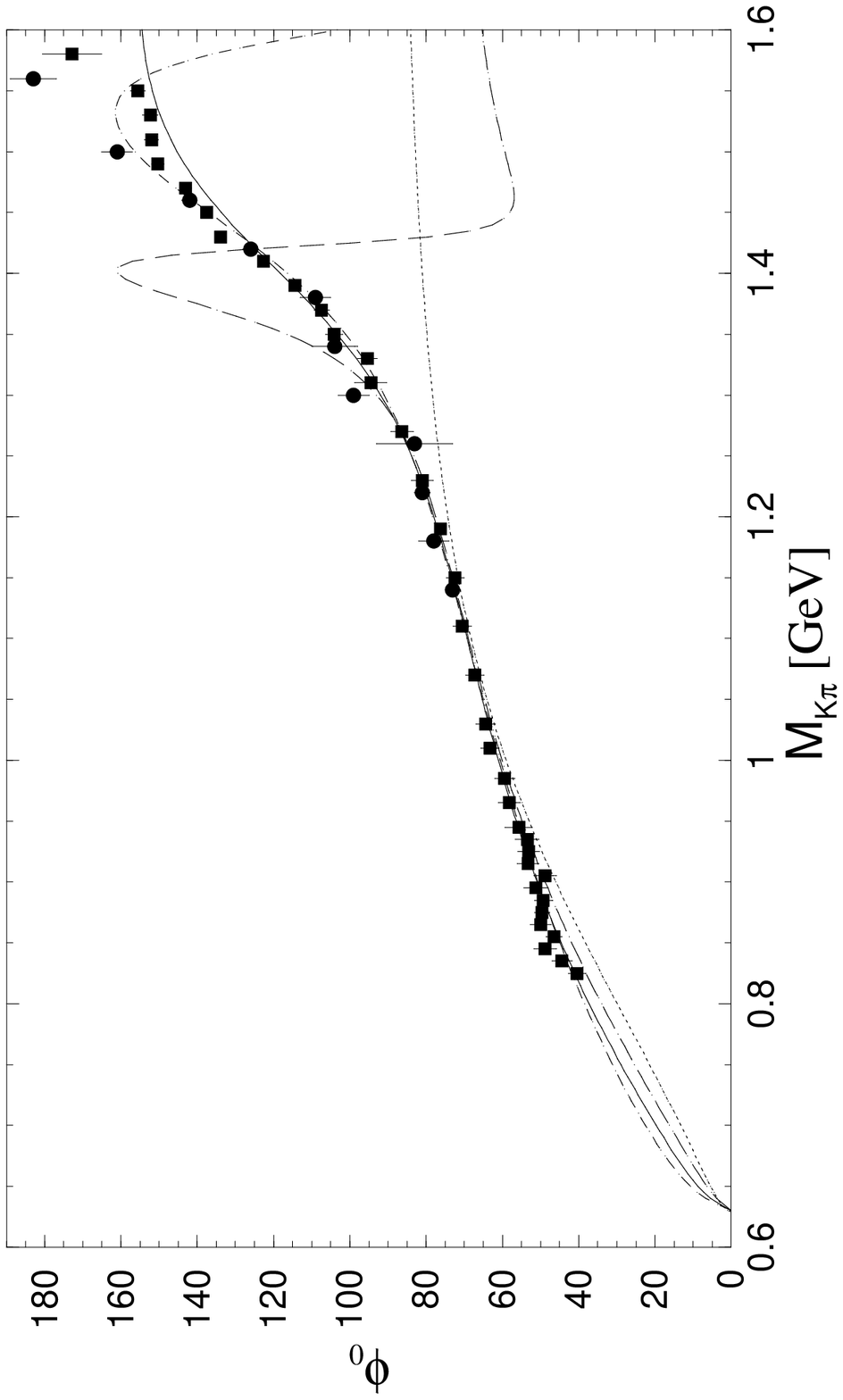} } }
\vspace{-4mm}
\caption[]{Elastic fits for $\phi_0$. Full circles: data of
ref.~\cite{est:78}; full squares: data of ref.~\cite{ast:88,awa:86}.
Long-dashed line: fit of eq.~\eqn{eq:4.6}; solid line: fit of
eq.~\eqn{eq:4.10}; short-dashed line: fit of eq.~\eqn{eq:4.13}; dotted
line: fit of eq.~\eqn{eq:4.14}.
\label{fig:3}}
\end{figure}
\begin{figure}[htb]
\centerline{
\rotate[r]{
\epsfysize=15cm
\epsffile{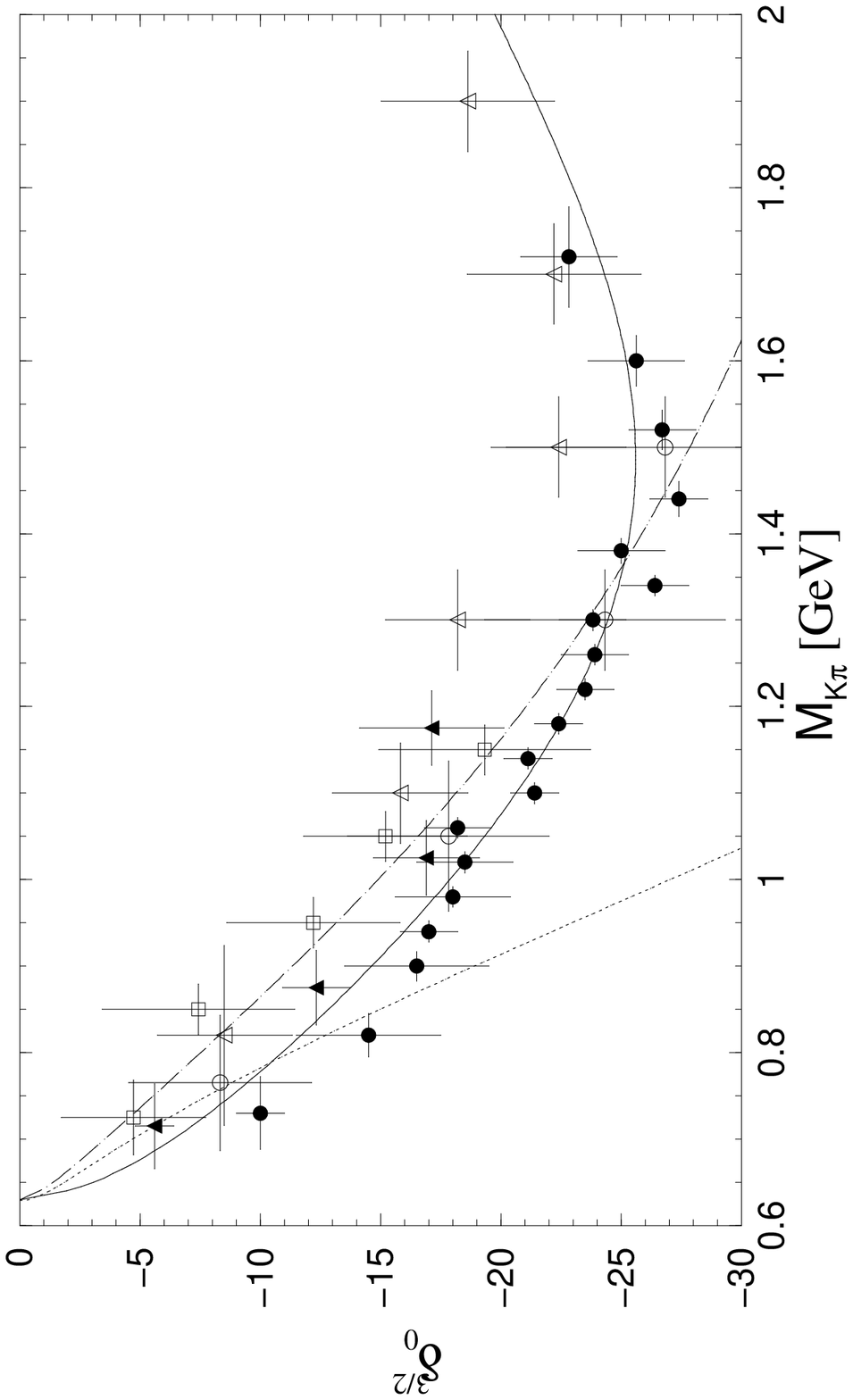} } }
\vspace{-4mm}
\caption[]{Elastic fits for $\delta_0^{3/2}$. The notation for the data sets
is as follows: ref. \cite{est:78} full circles; ref. \cite{jon:73}
full triangles; ref. \cite{lin:73} open circles; ref. \cite{bak:70} open
squares; ref. \cite{cho:70} open triangles. Long-dashed line: fit of
eq.~\eqn{eq:4.6}; solid line: fit of eq.~\eqn{eq:4.10}; dotted line: fit of
eq.~\eqn{eq:4.14}.
\label{fig:4}}
\end{figure}
The fit result corresponding to eq.~\eqn{eq:4.6} is shown as the long-dashed
line in figs.~1-4 for $\delta_0^{1/2}$, $a_0$, $\phi_0$ and $\delta_0^{3/2}$
respectively as a function of the invariant mass of the $K\pi$ system
$M_{K\pi}=\sqrt{s}$. The curves corresponding to the other two fits look
similar. In figs.~2 and 3, data for $a_0$ and $\phi_0$ up to $1.6\;\gev$
has been included already. The notation for the different data sets is as
follows: ref. \cite{est:78} full circles; ref. \cite{ast:88,awa:86} full
squares; ref. \cite{jon:73} full triangles; ref. \cite{lin:73} open circles;
ref. \cite{bak:70} open squares; ref. \cite{cho:70} open triangles. As
discussed above the errors for the data by refs.~\cite{est:78,jon:73,lin:73,
bak:70,cho:70} have been increased by a factor of two. Concerning the
quality of our fits we observe that the total $\chi^2/d.o.f.$ is around two.
This is due to the fact that the $\chi^2$ for $\delta_0^{3/2}$ individually
is almost three. The individual $\chi^2$ of $\delta_0^{1/2}$, $a_0$ and
$\phi_0$ are all close to one. Even removing the data for $\delta_0^{3/2}$
above $1.3\;\gev$ only reduces the $\chi^2$ for $\delta_0^{3/2}$ to about two.
Thus we conclude that the unitarised chiral expression does not provide a
fully satisfactory fit to the $\delta_0^{3/2}$ data and that it somewhat
prefers the older data. Nevertheless, compared to the conventional $\chi$PT
result without unitarisation \cite{bkm:91a,bkm:91b}, the situation is much
improved. For a comparison with the inverse amplitude method (IAM) see also
ref. \cite{dp:93}.

The somewhat unsatisfactory description of the $I=3/2$ data is probably due
to the fact that there is no resonance contribution in the s-channel. Being
pure background the chiral expressions are not expected to work very well at
higher energies, because additional contributions not included in our
low-energy description (e.g. higher-mass resonance exchanges in the $t$ and
$u$ channels) could be important numerically. This is reflected in the fact
that the contribution from unphysical cuts is much larger in the case of
$I=3/2$ than for $I=1/2$. However, the discrepancies among different sets of
data, and the fact that the older $I=3/2$ data admits a better theoretical
description, prevents us from drawing any clear conclusion. More precise
experimental data are needed in this channel.

To further improve our overall fit for all data sets, we decided to use a
unitarised background parametrisation for the $\delta_0^{3/2}$ data:
\begin{equation}
\label{eq:4.7}
\tan\delta_0^{3/2}(s) \;=\; \frac{\alpha\,q_{K\pi}(s)}{\Big(1+
\beta\,q_{K\pi}^2(s)+\gamma\,q_{K\pi}^4(s)\Big)} \,,
\end{equation}
where $q_{K\pi}(s)=\sqrt{s}\,\sigma_{K\pi}(s)/2$ is the $K\pi$ centre-of-mass
momentum and $\alpha$, $\beta$ and $\gamma$ are parameters of the ansatz.
The parameter $\alpha$ can be identified with the scattering length.
Again, we perform the same three fits as with the chiral expression for the
$I=3/2$ channel. First fixing $c_d$ and $c_m$ to the central values of
eq.~\eqn{eq:4.3}, we obtain
\begin{displaymath}
M_{K^*_0} \;=\; 1.28\;\gev \,, \qquad
c_{K\pi}^{1/2} \;=\; -\,0.130 \,,
\end{displaymath}
\begin{equation}
\label{eq:4.8}
\alpha \;=\; -\,0.897\;\gev^{-1} \,, \qquad
\beta \;=\; -\,0.284\;\gev^{-2} \,, \qquad
\gamma \;=\; 1.945\;\gev^{-4} \,,
\end{equation}
with a $\chi^2 = 107.7/115\,d.o.f. = 0.94$. The fit parameters imposing the
constraint $c_m=c_d$ are found to be
\begin{displaymath}
M_{K^*_0} \;=\; 1.19\;\gev \,, \qquad
c_d \;=\; c_m \;=\; 44.8\;\mev \,, \qquad
c_{K\pi}^{1/2} \;=\; -\,0.381 \,,
\end{displaymath}
\begin{equation}
\label{eq:4.9}
\alpha \;=\; -\,0.859\;\gev^{-1} \,, \qquad
\beta \;=\; -\,0.580\;\gev^{-2} \,, \qquad
\gamma \;=\; 2.301\;\gev^{-4} \,,
\end{equation}
and $\chi^2 = 96.4/114\,d.o.f. = 0.85$. Finally, leaving both $c_d$ and $c_m$
as free parameters yields
\begin{displaymath}
M_{K^*_0} \;=\; 1.36\;\gev \,, \qquad
c_d \;=\; 13.0\;\mev \,, \qquad
c_m \;=\; 85.4\;\mev \,, \qquad
c_{K\pi}^{1/2} \;=\; 0.200 \,,
\end{displaymath}
\begin{equation}
\label{eq:4.10}
\alpha \;=\; -\,0.860\;\gev^{-1} \,, \qquad
\beta \;=\; -\,0.567\;\gev^{-2} \,, \qquad
\gamma \;=\; 2.283\;\gev^{-4} \,,
\end{equation}
with $\chi^2 = 90.5/113\,d.o.f. = 0.80$.
The last fit is shown as the solid curve in figs.~1-4. Compared to the fits
with the chiral expression for the $I=3/2$ amplitude, the total $\chi^2/d.o.f.$
is greatly improved. The individual $\chi^2$ for $\delta_0^{1/2}$ is around
one and all other $\chi^2$ are below one. The parameters $\alpha$, $\beta$ and
$\gamma$ have now replaced the parameter $c_{K\pi}^{3/2}$ in the original
parametrisation of the $I=3/2$ amplitude. We observe that for the fits with
$c_d$ unequal $c_m$, the $K^*_0$ mass turns out to be somewhat larger than the
results in eq.~\eqn{eq:4.5} and \eqn{eq:4.9} and the value of $c_{K\pi}^{1/2}$
depends very much on the other parameters. However, again rather different
values of $c_d$ and $c_m$ lead to similar $\chi^2$ implying sizeable
uncertainties for these parameters.

Next, we compare our fits for the $K\pi$ scattering amplitudes with the
unitarised chiral expressions to a parametrisation for the $I=1/2$
amplitude with a $K$-matrix ansatz \cite{lan:75,lan:78,lp:80}, keeping
the background parametrisation of eq.~\eqn{eq:4.7} together with the
parameters of eq.~\eqn{eq:4.10} for the $I=3/2$ amplitude. The conventional
$K$-matrix ansatz is completely analogous to eq.~\eqn{eq:4.2},
\begin{equation}
\label{eq:4.11}
A_0^{1/2}(s) \;=\; \frac{\sigma_{K\pi}(s)K_0^{1/2}(s)}
{\Big(1-C(s)\,K_0^{1/2}(s)\Big)} \,,
\end{equation}
where now $C(s)=16\pi\bar J_{K\pi}(s)$, that is $c_{K\pi}^{1/2}=0$, and
\begin{equation}
\label{eq:4.12}
K_0^{1/2}(s) \; = \; \frac{g_R^2 M_R^2}{(M_R^2-s)} +
\frac{b\,(s-s_+)}{(1+c\,s)} \,.
\end{equation}
The $K$-matrix ansatz consists of a resonance term, with $g_R$ being the
coupling of the resonance, and a background contribution. We have tried
different forms of the background term, but the fit does not depend very
much on this choice. In figures~1-3 the $K$-matrix fit corresponds to
the short-dashed lines and the corresponding fit parameters take the values
\begin{displaymath}
M_R \;=\; 1.32\;\gev \,, \qquad
g_R \;=\; 0.628 \,,
\end{displaymath}
\begin{equation}
\label{eq:4.13}
b \;=\; 0.695\;\gev^{-2} \,, \qquad
c \;=\; 0.172\;\gev^{-2} \,.
\end{equation}
The $\chi^2 = 86.9/116\,d.o.f. = 0.75$ for this fit. Thus the fit quality
is very similar to our fit \eqn{eq:4.10} with the chiral expressions where
we also used a pure background form for the $I=3/2$ amplitude.

Nevertheless, as can already be anticipated from the curvature of the two
fits close to the $K\pi$ threshold, the $I=1/2$ scattering lengths turn
out to be somewhat different. Expanding our results close to threshold,
we obtain $\alpha_0^{1/2}(\mbox{K-matrix})=0.31\,M_\pi^{-1}$, whereas
$\alpha_0^{1/2}(\mbox{chiral})=0.18\,M_\pi^{-1}$, although the quality
of the fit for $I=1/2$ is equally good. As has become conventional in
the literature, we have expressed the scattering lengths in units of
$M_\pi^{-1}$. Even more dramatically, for the $I=3/2$ scattering lengths
we find $\alpha_0^{3/2}(\mbox{background})=-\,0.12\,M_\pi^{-1}$ and
$\alpha_0^{3/2}(\mbox{chiral})=-\,0.05\,M_\pi^{-1}$. Our $\chi$PT results
for the scattering lengths are in agreement with the order $p^4$ calculation
of ref.~\cite{bkm:91a} (table~3). These findings show that reasonable fits
to the experimental data can be obtained that spoil the $\chi$PT behaviour
in the low-energy region close to the $K\pi$ threshold. We conclude that the
scattering lengths for S-wave $K\pi$ scattering can only be extracted more
reliably once better experimental data close to threshold become available.

Finally, to get some feeling about the importance of the resonance
contribution and the real part of the unitarity corrections, as the dotted
line in figs.~1-4, we display the leading order $\chi$PT result using a
unitarisation according to the prescription:
\begin{equation}
\label{eq:4.14}
\tan\delta_0^I \; = \; \sigma_{K\pi}(s)\,t_0^{I\,(2)}(s) \,.
\end{equation}
This prescription is equivalent to eq.~\eqn{eq:3.1} with the choice
$g(s)=i\,\sigma_{K\pi}(s)$ and $N_0^I(s)=t_0^{I\,(2)}(s)$. We should remark
that this result is a parameter free prediction of lowest-order $\chi$PT.
However, figs.~1-4 show that, except for $\phi_0$ below roughly $1.2\;\gev$,
resonance and the real part of unitarity corrections are required in order
to obtain an acceptable description of the data.

\newsection{\boldmath Including the $K\eta$ and $K\eta'$ channels\unboldmath}

Above roughly $1.3\;\gev$, $K\pi$ scattering in the $I=1/2$ channel
was found to be inelastic \cite{est:78,ast:88} with $K\eta'$ being the first
important inelastic channel. Therefore, in order to be able to describe
$K\pi$ scattering beyond $1.3\;\gev$, we need to include the $K\eta'$
channel in the scattering amplitude in the framework of $\chi$PT with
resonances. Although the $K\eta$ channel was found to be unimportant,
nevertheless, for consistency we have also included it in our analysis.
The importance of the $K\eta$ channel for our fits shall be further
discussed below.

In the large-$N_c$ limit, the singlet $\eta_1$ field becomes the ninth
Goldstone boson and can then be incorporated with an extended
$U(3)_L \otimes U(3)_R$ chiral Lagrangian \cite{her:97,her:98}. The large mass
of the $\eta'$ originates from the $U(1)_A$ anomaly, which, although formally
of ${\cal O}(1/N_c)$, is numerically important and cannot be treated as a
small perturbation. It is possible to build a consistent combined chiral
expansion in powers of momenta, quark masses and $1/N_c$, by counting the
relative magnitude of these parameters as
$m_q \sim 1/N_c \sim {\cal O}(p^2)$ \cite{leu:96}.

The physical states $\eta$ and $\eta'$ are mixtures of the SU(3)
singlet and octet states $\eta_1$ and $\eta_8$. If isospin-breaking effects
are neglected, the mixing can be parametrised in terms of the mixing angle
$\theta$ usually defined by

\newpage

\begin{eqnarray}
\label{eq:5.1}
|\eta\rangle  &=& \cos\theta\,|\eta_8\rangle-\sin\theta\,|\eta_1\rangle\,,\nn\\
|\eta'\rangle &=& \sin\theta\,|\eta_8\rangle + \cos\theta\,|\eta_1\rangle \,.
\end{eqnarray}
For our numerical analysis below, we shall use $\sin\theta=-1/3\approx-\,
20^\circ$ which is in the ball-park of present day phenomenological values.
Further details on $\eta$-$\eta'$ mixing in the framework of $\chi$PT
with resonances can be found in appendix~B. To be able to include the
$K\eta$ and  $K\eta'$ channels in the analysis of $K\pi$ scattering we have
to calculate the five scattering amplitudes $T_{ K\pi\to K\eta_i}$ and
$T_{ K\eta_i\to K\eta_j}$ with $i,j=1,8$. The resulting expressions are rather
lengthy and therefore, although representing one of our main results, have
been relegated to appendix~C. Eqs.~\eqn{eq:C1} to \eqn{eq:C5} include the
leading order $p^2$ and the resonance contributions. So far, we have not
calculated the order $p^4$ loop contributions to the inelastic channels,
but as will be explained in more detail below, the dominant contributions
are accounted for by our unitarisation procedure. Using eq.~\eqn{eq:5.1},
the five required amplitudes $T_{K\pi\to K\eta}$, $T_{K\pi\to K\eta'}$,
$T_{K\eta\to K\eta}$, $T_{K\eta\to K\eta'}$ and $T_{K\eta'\to K\eta'}$ can
then easily be expressed in terms of $T_{ K\pi\to K\eta_i}$ and
$T_{ K\eta_i\to K\eta_j}$.

For a coupled channel analysis, our unitarisation procedure of section~3
has to be generalised and in particular eq.~\eqn{eq:3.1} has to be modified
to a matrix equation. In the three-channel case, $N(s)$ and $g(s)$ become
three by three matrices. Since in the following only S-wave $K\pi$ scattering
in the $I=1/2$ channel is considered, to simplify the notation we have
dropped the corresponding indices in the expressions. Then the generalisation
of eq.~\eqn{eq:3.1} takes the form \cite{oo:99,oor:00}
\begin{equation}
\label{eq:5.3}
\tilde t(s) \;=\; N(s)\,\Big[\,1-g(s)\,N(s)\,\Big]^{-1} \,,
\end{equation}
with
\begin{equation}
\label{eq:5.4}
N(s) \;=\; \left(\!\! \begin{array}{ccc}
N_{K\pi \to K\pi  }(s) & N_{K\pi \to K\eta }(s) & N_{K\pi  \to K\eta'}(s) \\
N_{K\pi \to K\eta }(s) & N_{K\eta\to K\eta }(s) & N_{K\eta \to K\eta'}(s) \\
N_{K\pi \to K\eta'}(s) & N_{K\eta\to K\eta'}(s) & N_{K\eta'\to K\eta'}(s)
\end{array} \!\!\right) \,,
\end{equation}
and
\begin{equation}
\label{eq:5.5}
g(s) \;=\; \left(\!\! \begin{array}{ccc}
g_{K\pi}(s) &            0 &             0 \\
          0 & g_{K\eta}(s) &             0 \\
          0 &            0 & g_{K\eta'}(s) \end{array} \!\!\right) \,.
\end{equation}
Several remarks are in order at this point. Because of time-reversal
invariance, $N(s)$ has to be symmetric. In addition, to satisfy unitarity
$N_{ij}^{-1}(s)$ needs to be real above the highest threshold of the channels
$i$ and $j$, and
\begin{equation}
\label{eq:5.6}
\IM\, g_i(s) \;=\; \theta(s-s_i)\,\sigma_i(s) \,,
\end{equation}
where $i$ denotes the three channels $K\pi$, $K\eta$ or $K\eta'$, $s_i$ is
the corresponding threshold, and the $\sigma_i(s)$ are defined in analogy
to eq.~\eqn{eq:3.5}.

$N_{K\pi\to K\pi}$ is already given by eq.~\eqn{eq:3.2}, and
$N_{K\pi\to K\eta}$, $N_{K\pi\to K\eta'}$, $N_{K\eta\to K\eta}$,
$N_{K\eta\to K\eta'}$ as well as $N_{K\eta'\to K\eta'}$ result from the
S-wave projection of the corresponding $T$ amplitudes according to eq.
\eqn{eq:2.8}. No subtractions have to be performed for these $N(s)$ because
we have not included the loop contributions to the $K\eta$ and $K\eta'$
channels explicitly. Therefore in this case $N(s)$ is actually equal to
$t(s)$ (compare with eq.~\eqn{eq:3.2}). Finally, analogous to eq.~\eqn{eq:3.3},
$g_{K\eta}(s)$ and $g_{K\eta'}(s)$ are written as
\begin{eqnarray}
\label{eq:5.7}
g_{K\eta }(s)  & = & 16\pi\bar J_{K\eta }(s) + c_{K\eta } \,, \nn \\
g_{K\eta'}(s)  & = & 16\pi\bar J_{K\eta'}(s) + c_{K\eta'} \,,
\end{eqnarray}
which satisfies eq.~\eqn{eq:5.6} automatically. Again, $c_{K\eta }$ and
$c_{K\eta'}$ are arbitrary real constants.

As it stands, eq.~\eqn{eq:5.3} already includes the dominant part of the
unitarity corrections which arise from the loop diagrams in $\chi$PT also
for the inelastic channel. The imaginary parts are generated by the imaginary
parts of $g_{K\eta}(s)$ and $g_{K\eta'}(s)$. We have verified explicitely
that the imaginary part generated from the leading contribution to the
$T_{K\pi\to K\eta_8}$ amplitude agrees with the imaginary part of the
$T_{K\pi\to K\pi}$ resulting from the $\eta_8$ terms of eq.~\eqn{eq:2.5}.
Thus, when using eq.~\eqn{eq:2.5} in our numerical analysis, for consistency
we have dropped the $\eta_8$ contribution since it arises from the
$T_{K\pi\to K\eta_8}$ amplitude. On the other hand, the real parts of the
unitarity corrections in the inelastic channels are not complete; the missing
pieces are parametrised by the constants $c_{K\eta}$ and $c_{K\eta'}$.

\newsection{Fitting the inelastic case}

For the inelastic case, the unitarised chiral expression corresponding to
the S-wave $I=1/2$ amplitude $A_0^{1/2}$ takes the form
\begin{equation}
\label{eq:6.1}
A_0^{1/2}(s) \;=\; \sqrt{\sigma(s)}\, N(s)\,\Big[\,1-g(s)\,N(s)\,\Big]^{-1}
\sqrt{\sigma(s)} \,,
\end{equation}
where
\begin{equation}
\label{eq:6.2}
\sigma(s) \;=\; \left(\!\! \begin{array}{ccc}
\sigma_{K\pi}(s) &                 0 &                  0 \\
               0 & \sigma_{K\eta}(s) &                  0 \\
               0 &                 0 & \sigma_{K\eta'}(s)
\end{array} \!\!\right) \,.
\end{equation}
The $K\pi\to K\pi$ contribution is obtained by taking the relevant
matrix element of the matrix $A_0^{1/2}(s)$. If the $K\eta$ and $K\eta'$
contributions are removed one of course recovers the corresponding expression
\eqn{eq:4.2} for the single channel case.


Let us next discuss the situation of the experimental data in the inelastic
region above $1.3\;\gev$. Because experimentally only the absolute value of
the scattering amplitude can be determined, there are $2^L$ discrete
ambiguous solutions to the partial wave decomposition of the data where
$L$ is the maximal angular momentum that contributes at a given energy.
These discrete ambiguities are apparent from the behaviour of the imaginary
part of the amplitude (Barrelet) zeros $z_i$\cite{bar:72}, if the scattering
amplitude is expressed as a complex polynomial in $z=\cos\theta$ where
$\theta$ is the scattering angle:
\begin{equation}
\label{eq:6.5}
A(s,z) \; = \; \sum\limits_{l=0}^L\, (2l+1) A_l(s) P_l(z) 
\; = \; f(s) \prod\limits_{i=1}^L\, (z-z_i(s)) \,.
\end{equation}
The measurement of $|A(s,z)|$ cannot determine whether $z_i(s)$ or $z_i^*(s)$
is a zero of $A(s,z)$; thus the $2^L$ ambiguity arises. Requiring the solutions
to be smooth, it is possible to switch from one solution to another when a
Barrelet zero $z_i$ approaches the real axis. In the elastic region the
ambiguity can be fixed from unitarity. However, in the inelastic region the
behaviour of the Barrelet zeros has to be studied in detail.

Whereas in the experiment by Estabrooks et al. \cite{est:78} a fourfold
ambiguity arose above $1.5\;\gev$, because the imaginary parts of the two
Barrelet zeros $z_1$ and $z_2$ were found to vanish within the experimental
uncertainties, the group by Aston et al. \cite{ast:88} obtained an unambiguous
solution up to $1.85\;\gev$. In the region between $1.86\;\gev$ and $2.0\;\gev$
Aston et al. found two solutions whereas above $2.0\;\gev$ four solutions
to the partial wave decomposition remained. Inspecting the signs of the
imaginary parts of the $z_i$ it is clear that the solution of Aston et al.
below $1.85\;\gev$ corresponds to solution~B of Estabrooks et al. Since both
experiments have similar statistics in our opinion the question whether the
solution of Aston et al. in the region between $1.5\;\gev$ and $1.85\;\gev$
is the physical one, requires further corroboration. Nevertheless, we have
decided to perform our fits to the experimental data by Aston et al.
\cite{ast:88,awa:86} and solution B of Estabrooks et al. \cite{est:78} up
to $1.85\;\gev$ with the unitarised chiral expressions.

In addition, we have also performed chiral fits to the other solutions of
Estabrooks et al. \cite{est:78} above $1.5\;\gev$. It is found that fits with
reasonable values for the parameters can be obtained for data sets A, B and D
whereas we were not able to find an acceptable fit for data set C. As far as
the $\chi^2$ is concerned, the best fit was obtained for data set A with
$\chi^2=1.70$, the next best was data set B with $\chi^2=1.86$ and finally
data set D with $\chi^2=2.57$. Therefore, on the basis of these fits, we are
unable to confirm that the solution found by the group of Aston et al.
\cite{ast:88} is indeed the physical one. Note, however, that in
ref.~\cite{ls:78} arguments were put forward which entail that the solutions
A and D of \cite{est:78} are unphysical. This result, together with our fits,
then favours solution B, in agreement with the findings of ref.~\cite{ast:88}.
In what follows, we shall not discuss the additional solutions of Estabrooks
et al. any further.

In the region of $1.9\;\gev$, the experiment by Aston et al. found a
second scalar resonance in the S-wave channel. If we aim to describe the
data up to such energies a second scalar nonet should be implemented in
the unitarised chiral expressions. To this end we add in eqs.~\eqn{eq:2.4}
and \eqn{eq:C1} to \eqn{eq:C5} a second nonet with degenerate mass $M_{S'}$
and new scalar couplings $c_d'$ and $c_m'$ analogous to the first nonet.
In addition we have also included two more vector octets with new couplings
$G_{V'}$ and $G_{V''}$. These couplings can be estimated from the decay rates
$K^*(1410)\to K\pi$ and $K^*(1650)\to K\pi$ \cite{pdg:98} with the results
$G_{V'}=12\;\mev$ and $G_{V''}=23\;\mev$, presumably with large errors. For
our first fit as in section~4 we fix the values of $c_d$ and $c_m$ to the
central values of eq.~\eqn{eq:4.3}. Since we would like to have a good
description for the $I=3/2$ channel up to $1.9\;\gev$, we chose to employ
the unitarised background expression \eqn{eq:4.7} as described in section~4.
Using the chiral expressions also for $I=3/2$, even including the higher
scalar and vector resonances, did not provide acceptable fits in the region
above $1.5\;\gev$. The fit parameters are then found to be
\begin{displaymath}
M_{K^*_0} \;=\; 1.26\;\gev \,, \qquad
M_{S'} \;=\; 2.17\;\gev \,, \qquad
c_d' \;=\; 31.6\;\mev \,, \qquad
c_m' \;=\; 46.9\;\mev \,,
\end{displaymath}
\begin{equation}
\label{eq:6.6}
c_{K\pi}^{1/2} \;=\; 0.174 \,, \qquad
c_{K\eta} \;=\; -\,0.890 \,, \qquad
c_{K\eta'} \;=\; -\,0.722 \,,
\end{equation}
\vspace{-4mm}
\begin{displaymath}
\alpha \;=\; -\,0.760\;\gev^{-1} \,, \qquad
\beta \;=\; -\,1.300\;\gev^{-2} \,, \qquad
\gamma \;=\; 3.373\;\gev^{-4} \,,
\end{displaymath}
with a $\chi^2 = 753.3/166\,d.o.f. = 4.54$. The resulting $\chi^2$ is found
to be rather large indicating that the estimate of eq.~\eqn{eq:4.3} for $c_d$
and $c_m$ does not provide a good description of the data if inelastic
channels and a second scalar resonance are included. We shall come back to
a discussion of this point below. Analogous to section~4, the next fit is
performed with the constraints $c_m=c_d$ and $c_m'=c_d'$. The resulting fit
parameters are
\begin{displaymath}
M_{K^*_0} \;=\; 1.29\;\gev \,, \qquad
M_{S'} \;=\; 1.86\;\gev \,, \qquad
c_d  \;=\; 25.8\;\mev \,, \qquad
c_d' \;=\; 14.8\;\mev \,,
\end{displaymath}
\begin{equation}
\label{eq:6.7}
c_{K\pi}^{1/2} \;=\; 0.179 \,, \qquad
c_{K\eta} \;=\; 0.504 \,, \qquad
c_{K\eta'} \;=\; -\,1.064 \,,
\end{equation}
\vspace{-4mm}
\begin{displaymath}
\alpha \;=\; -\,0.725\;\gev^{-1} \,, \qquad
\beta \;=\; -\,1.817\;\gev^{-2} \,, \qquad
\gamma \;=\; 4.385\;\gev^{-4} \,,
\end{displaymath}
with a $\chi^2 = 470.9/166\,d.o.f. = 2.84$. This fit is shown as the
short-dashed line in figs.~5 and 6 for the modulus of the amplitude $a_0$ and
the phase $\phi_0$ respectively. For comparison, in addition, as the dotted
line we display the best fit for the elastic case of section~4,
eq.~\eqn{eq:4.10}.
\begin{figure}[htb]
\centerline{
\rotate[r]{
\epsfysize=15cm
\epsffile{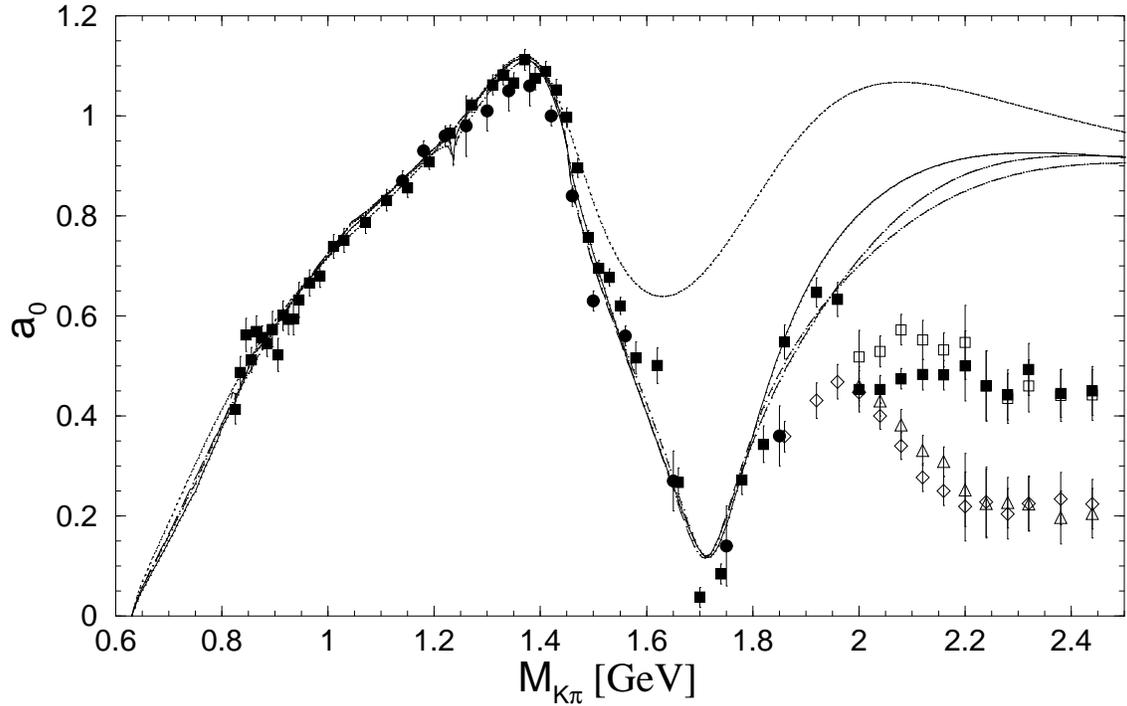} } }
\vspace{-4mm}
\caption[]{Inelastic fits for $a_0$. The notation for the data sets is as
follows: solution B of \cite{est:78} full circles; solution A of \cite{ast:88,
awa:86} full squares; solution B of \cite{ast:88,awa:86} empty diamonds;
solution C of \cite{ast:88,awa:86} empty squares; solution D of \cite{ast:88,
awa:86} empty triangles. Dotted line: fit of eq.~\eqn{eq:4.10}; short-dashed
line: fit of eq.~\eqn{eq:6.7}; solid line: fit of eq.~\eqn{eq:6.8}; long-dashed
line:  fit of eq.~\eqn{eq:6.11}.
\label{fig:5}}
\end{figure}
\begin{figure}[htb]
\centerline{
\rotate[r]{
\epsfysize=15cm
\epsffile{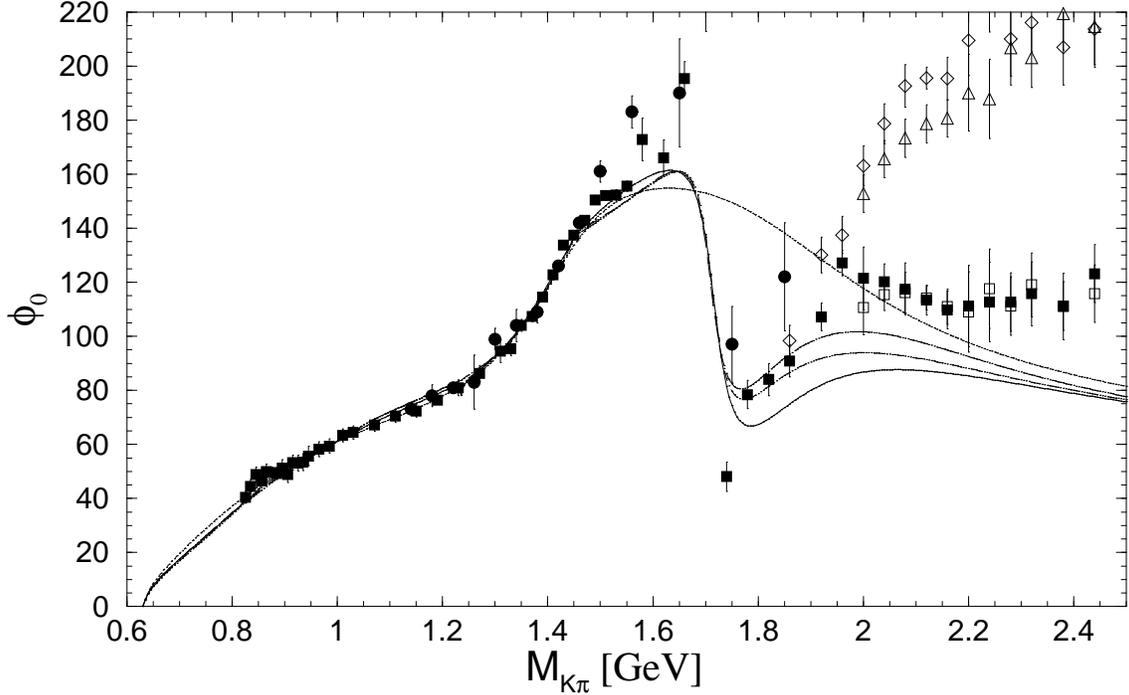} } }
\vspace{-4mm}
\caption[]{Inelastic fits for $\phi_0$. The notation for the data sets is as
follows: solution B of \cite{est:78} full circles; solution A of \cite{ast:88,
awa:86} full squares; solution B of \cite{ast:88,awa:86} empty diamonds;
solution C of \cite{ast:88,awa:86} empty squares; solution D of \cite{ast:88,
awa:86} empty triangles. Dotted line: fit of eq.~\eqn{eq:4.10}; short-dashed
line: fit of eq.~\eqn{eq:6.7}; solid line: fit of eq.~\eqn{eq:6.8}; long-dashed
line:  fit of eq.~\eqn{eq:6.11}.
\label{fig:6}}
\end{figure}

Our notation for the shown data sets is as follows: solution B of Estabrooks
et al. \cite{est:78} full circles; solution A of Aston et al. \cite{ast:88,
awa:86} full squares; solution B of \cite{ast:88,awa:86} empty diamonds;
solution C of \cite{ast:88,awa:86} empty squares; solution D of \cite{ast:88,
awa:86} empty triangles. Although in our fits we have only included data of
Estabrooks et al. and Aston et al. up to $1.85\;\gev$ where the partial wave
analysis of Aston et al. gave a unique solution, in our figures all data up to
$2.5\;\gev$ are displayed for convenience. Data points where $\phi$ is larger
than $180^\circ$ have not been included in the fit since they violate
unitarity. This is easily seen from the relation
\begin{equation}
\label{eq:6.7a}
a_0\sin(\phi_0) \;=\;
\frac{1}{2}\,\biggl( 1 - \eta_0^{1/2}\cos(2\delta_0^{1/2}) \biggr) +
\frac{1}{4}\,\biggl( 1 - \eta_0^{3/2}\cos(2\delta_0^{3/2}) \biggr) \,,
\end{equation}
which can be derived immediately by taking the imaginary part of
eq.~\eqn{eq:4.1}, because for $0\leq\eta_0^I\leq 1$ the right-hand side is
always positive. This shows that solutions B and D of \cite{ast:88,awa:86},
also displayed in figs.~5 and 6, are unphysical.

Finally, we perform a fit leaving both $c_d$ and $c_m$ as free parameters
but keeping the constraint $c_m'=c_d'$ because otherwise we would have too
many parameters, and the fit would become unstable:
\begin{displaymath}
M_{K^*_0} \;=\; 1.30\;\gev \,, \qquad
M_{S'} \;=\; 1.84\;\gev \,, \qquad
c_d \;=\; 29.3\;\mev \,, \qquad
c_m \;=\;  8.8\;\mev \,,
\end{displaymath}
\begin{equation}
\label{eq:6.8}
c_d' \;=\; 13.1\;\mev \,, \qquad
c_{K\pi}^{1/2} \;=\; 0.151 \,, \qquad
c_{K\eta} \;=\; 1.369 \,, \qquad
c_{K\eta'} \;=\; -\,1.388 \,,
\end{equation}
\vspace{-4mm}
\begin{displaymath}
\alpha \;=\; -\,0.816\;\gev^{-1} \,, \qquad
\beta \;=\; -\,1.045\;\gev^{-2} \,, \qquad
\gamma \;=\; 3.436\;\gev^{-4} \,,
\end{displaymath}
with a $\chi^2 = 408.8/165\,d.o.f. = 2.48$. The resulting amplitude and phase
are shown as the solid line in figs.~5 and 6.

As for the fits in the elastic region presented in section~4, also in the
inelastic case rather different values for the scalar couplings can lead
to similar fit quality. Therefore, as in section~4, we next try to put further
constraints on these couplings to single out physically sensible values.
As will be discussed in a subsequent publication \cite{jop:00b}, the
high-energy or short-distance behaviour of the scalar form factor can provide
such constraints. Requiring the tree-level form factor in the large-$N_c$ limit
to vanish at infinite momentum transfer, for an arbitrary number of scalar
resonances, one finds the following relations:
\begin{equation}
\label{eq:6.9}
4\sum\limits_i c_{mi}\,c_{di} \; = \; f_\pi^2
\qquad \mbox{and} \qquad
\sum\limits_i \frac{c_{mi}}{M_{S_i}^2}\,(c_{mi}-c_{di}) \; = \; 0 \,,
\end{equation}
where $M_{S_i}$, $c_{di}$ and $c_{mi}$ are the mass and the couplings of
the $i$th resonance respectively.

In the case of two resonances, the short-distance constraints of
eq.~\eqn{eq:6.9} can be used to express the couplings of the second resonance
$c_d'$ and $c_m'$ in terms of the resonance masses and the couplings of the
first resonance. For this fit, we find that $c_{K\eta}$ is zero within the
fit uncertainties. Thus we decided to set $c_{K\eta}=0$. Fitting the remaining
parameters, we obtain
\begin{displaymath}
M_{K^*_0} \;=\; 1.41\;\gev \,, \qquad
M_{S'} \;=\; 2.69\;\gev \,, \qquad
c_d \;=\; 22.8\;\mev \,,
\end{displaymath}
\begin{equation}
\label{eq:6.10}
c_m \;=\; 26.0\;\mev \,, \quad
c_{K\pi}^{1/2} \;=\; 0.178 \,, \quad
c_{K\eta'} \;=\; 0.151 \,,
\end{equation}
\vspace{-4mm}
\begin{displaymath}
\alpha \;=\; -\,0.750\;\gev^{-1} \,, \qquad
\beta \;=\; -\,1.553\;\gev^{-2} \,, \qquad
\gamma \;=\; 4.041\;\gev^{-4} \,,
\end{displaymath}
with a $\chi^2 = 505.6/167\,d.o.f. = 3.03$. The values for $c_d'$ and $c_m'$
which resulted from the fit of eq.~\eqn{eq:6.10} are $c_d'=43.8\;\mev$ and
$c_m'=35.3\;\mev$. As our last fit, we employ the constraints \eqn{eq:6.9}
together with the reasonable assumption that $c_m=c_d$ and again $c_{K\eta}=0$.
In this case, the fit parameters are found to be
\begin{displaymath}
M_{K^*_0} \;=\; 1.40\;\gev \,, \qquad
M_{S'} \;=\; 2.57\;\gev \,, \qquad
c_d \;=\; 23.8\;\mev \,,
\end{displaymath}
\begin{equation}
\label{eq:6.11}
c_{K\pi}^{1/2} \;=\; 0.173 \,, \qquad
c_{K\eta'} \;=\; 0.112 \,,
\end{equation}
\vspace{-4mm}
\begin{displaymath}
\alpha \;=\; -\,0.758\;\gev^{-1} \,, \qquad
\beta \;=\; -\,1.473\;\gev^{-2} \,, \qquad
\gamma \;=\; 3.929\;\gev^{-4} \,,
\end{displaymath}
with a $\chi^2 = 507.1/168\,d.o.f. = 3.02$. The resulting $\chi^2$ is even
slightly smaller than for the previous fit without the additional constraint
$c_m=c_d$. For the couplings of the second resonance from this fit we obtain
$c_d'=c_m'=39.6\;\mev$. The latter fit is shown as the long-dashed line in
figs.~5 and 6. Let us now come to a general discussion about the quality of
the fits for the inelastic $K\pi$ scattering amplitude.

Even for our best fit of eq.~\eqn{eq:6.8}, the resulting $\chi^2$ is larger
than two. In our opinion the reason for this is twofold. On the one hand,
the systematic uncertainties for the data above $1.3\;\gev$ seem to be
underestimated. This is seen if the data are fit with a Breit-Wigner-resonance
plus background ansatz \cite{ast:88}. Even increasing the number of
parameters in the parametrisation of the background, the lowest $\chi^2$ is
found to be of the order of $1.5$. On the other hand, additional inelastic
channels like $K\pi\pi\pi$ are still missing in our theoretical expressions.
This could also lead to deficiencies in the description of the data in the
higher-energy region. However, in view of our first remark, the experimental
data are described rather well which in turn implies that additional inelastic
channels should not be too important. This is supported by the fact that
those contributions are of higher order in the chiral and $1/N_c$ expansions
and should be suppressed.

Comparing the various fits for the inelastic channel, like in section~4 one
finds that similar fit quality can be achieved with rather different values
for the couplings of the scalar resonances. In view of this fact it seemed
desirable to put further constraints on these couplings from general
considerations. These are provided by the short-distance constraints
\eqn{eq:6.9} which arise if one requires that the tree-level scalar form factor
in the large-$N_c$ limit should vanish at infinite momentum transfer. For this
reason, we believe that the fits of eqs.~\eqn{eq:6.7} and especially
\eqn{eq:6.11} are physically most sensible, although for the description of
the data, the fits of eqs. \eqn{eq:6.8} and \eqn{eq:6.10} lead to very similar
results.

As has been mentioned at the beginning of section~5, experimentally it was
observed that the $K\eta$ channel is not important for S-wave $K\pi$
scattering up to $2\;\gev$. Thus, to get a further understanding about which
of our fits is most physical, let us investigate how sensitive the fits are
concerning a removal of the $K\eta$ channel. For simplicity, we just consider
the fits of eqs.~\eqn{eq:6.7} and \eqn{eq:6.11}.

If the $K\eta$ channel is removed in the fit of eq.~\eqn{eq:6.7}, the $\chi^2$
is increased to $\chi^2 = 664/167\,d.o.f. = 3.98$ On the other hand, if an analogous
fit without the $K\eta$ channel and the constraints $c_m=c_d$ and $c_m'=c_d'$ 
is performed, a fit with similar quality than \eqn{eq:6.7} can be obtained:
\begin{displaymath}
M_{K^*_0} \;=\; 1.29\;\gev \,, \qquad
M_{S'} \;=\; 1.94\;\gev \,, \qquad
c_d  \;=\; 27.9\;\mev \,,
\end{displaymath}
\begin{equation}
\label{eq:6.14}
c_d' \;=\; 20.1\;\mev \,, \qquad
c_{K\pi}^{1/2} \;=\; 0.173 \,, \qquad
c_{K\eta'} \;=\; -\,0.860 \,,
\end{equation}
\vspace{-4mm}
\begin{displaymath}
\alpha \;=\; -\,0.846\;\gev^{-1} \,, \qquad
\beta \;=\; -\,0.917\;\gev^{-2} \,, \qquad
\gamma \;=\; 3.403\;\gev^{-4} \,,
\end{displaymath}
with a $\chi^2 = 479.6/167\,d.o.f. = 2.87$. Putting back the $K\eta$ channel
for these fit parameters only increases the $\chi^2$ to $522/167\,d.o.f. =
3.13$. Therefore, as far as the sensitivity with respect to the $K\eta$
channel is concerned, the fit \eqn{eq:6.14} is preferred over the fit
\eqn{eq:6.7}.

Performing the same exercise of removing the $K\eta$ channel in the fit of
eq.~\eqn{eq:6.11}, the $\chi^2$ is even slightly decreased to
$488/168\,d.o.f. = 2.90$. In this sense there is some indication that the
parameter set of eq. \eqn{eq:6.11} is the most physical of our fits to the
S-wave $K\pi$ scattering amplitude.

In the original publications where the inclusion of resonances in the
framework of $\chi$PT was worked out \cite{egpr:89,eglpr:89}, it was
suggested that the order $p^4$ coupling constants $L_5$ and $L_8$ are
saturated by the contribution of the scalar octet resonances to a very
good approximation. In the case of two resonances these contributions are
given by:
\begin{equation}
\label{eq:6.12}
L_5^S \;=\; \frac{c_d c_m}{M_S^2} + \frac{c_d' c_m'}{M_{S'}^2}
\qquad \mbox{and} \qquad
L_8^S \;=\; \frac{c_m^2}{2M_S^2} + \frac{c_m'^2}{2M_{S'}^2} \,.
\end{equation}
Thus, assuming resonance saturation of $L_5$ and $L_8$ and using our fit
parameters we are in a position to calculate these chiral couplings. Taking
the fit of eq.~\eqn{eq:6.11} including the short-distance constraint as an
example, we then find
\begin{equation}
\label{eq:6.13}
L_5^S \; = \; 0.52\cdot 10^{-3}
\qquad \mbox{and} \qquad
L_8^S \; = \; 0.26\cdot 10^{-3} \,,
\end{equation}
in good agreement to the most recent determination of these constants
\cite{abt:99,abt:00} and also supporting lower values than were found in
earlier estimates. The fact that in this approach we obtain $L_5=2L_8$ follows
immediately from the second constraint of eq.~\eqn{eq:6.9}. Similar values
for $L_5$ and $L_8$ are also obtained for the other fits to the inelastic
scattering amplitude.

\newsection{The $T$-matrix poles}

In the energy region above the $K\pi$ threshold and below roughly $2\;\gev$,
the possible presence of three scalar resonances in the S-wave $I=1/2$ channel
has been discussed in the literature. The only one which is clearly observed in
the experimental data is the $K^*_0(1430)$ \cite{est:78,ast:88}. Less obvious
is the existence of the $K^*_0(1950)$ resonance \cite{ast:88} in the region
around $2\;\gev$, and concerning the third, the lowest-lying $\kappa$ meson
\cite{jaf:77,sca:82,bev:86,iiitt:97,ds:98,bfss:98,oop:99,oo:99}, there has
been some controversy in recent years \cite{tor:95,as:97,cp:00}.

We now address this problem by searching for possible poles of the matrix
$\tilde t(s)$, eq.~\eqn{eq:5.3}, associated with
these resonances in the nearby unphysical Riemann sheets. As discussed above,
the influence of the $K\eta$ threshold in the S-wave $I=1/2$ partial wave
amplitude was found to be very small and hence, in order to simplify the
notation, we shall label the different Riemann sheets as if we had only a two
channel problem with $K\pi$ and $K\eta'$. In this way, the first or physical
Riemann sheet is indicated by $(+,+)$, the second one by $(-,+)$, the third
one by $(-,-)$ and the fourth by $(+,-)$. The signs between brackets refer to
the sign ambiguity present in the definition of the modulus of the
centre-of-mass three-momentum due to the presence of the square root of eq.
\eqn{eq:3.5}. In this way, the `$+$' sign indicates that the imaginary part of
the modulus of the three-momentum is always positive in the whole $s$-plane
and the `$-$' sign refers to the contrary. The sign between brackets on the
left side corresponds to the $K\pi$ system and the one on the right side
to the $K\eta'$ state. We start by investigating the pole positions for our
preferred fit of eq.~\eqn{eq:6.11}. At the end of this section, we shall also
compare with the pole positions found for some of the other fits obtained in
sections 4 and~6.

The nearby Riemann sheet for the $\kappa$ resonance is the second one, since
this state should appear with a mass much lower than the threshold of the
$K\eta'$ channel. In fact, we find a pole in the second Riemann sheet $(-,+)$
at
\begin{equation}
\label{eq:7.1}
\sqrt{s}_{\kappa} \:=\; (708-i\,305) \; \mev \,.
\end{equation}
This pole position is very similar to the one found in ref.~\cite{oo:99} at
$(779-i\,330)\;\mev$.\footnote{Employing the pole position one can identify
the real part with the mass of the resonance and the modulus of the imaginary
part with one half of its width. For narrow objects, this identification agrees
with the Breit-Wigner resonance picture.} Note that the main differences
between ref. \cite{oo:99}, applied to the specific case of S-wave $I=1/2$
scattering, and the present work are that in the former unphysical cut
contributions were not included, the $K\eta'$ channel was not considered
(although its influence below $1.3\;\gev$ is very smooth), and the short
distance constraints were not imposed. However, in the present work, although
we have included all this information, the pole \eqn{eq:7.1} corresponding to
the $\kappa$ resonance appears in a similar position to the one found in
ref. \cite{oo:99}. (We are talking about a resonance with a width of about
$600\;\mev$; hence small variations in its mass and width under realistic
changes in the dynamical input are completely natural.) These findings are
expected since in ref. \cite{oo:99} it was estimated that the contribution
from the unphysical cuts in the physical region below $0.8\;\gev$ amounts
just to a few percent. Thus the stability of the expansion in the physical
region in terms of the unphysical cut contributions at these energies seems
to be justified.

We have checked that this pole is of dynamical origin, that is, it is
generated through the strong rescattering (unitarity loops) of the $K\pi$
channel. For instance, this pole survives when one removes all the explicit
tree-level resonant contributions, as was already established in ref.
\cite{oo:99}. Because of this fact and also because the $\kappa$ is a very
wide resonance, which implies that its contribution to the physical data is
rather soft, one can expect that the existence and properties of this pole
are rather model dependent. This is basically what is observed in the recent
work \cite{cp:00}, where a different conclusion has been obtained. However,
contrary to what is concluded in this reference, the model dependence does
not imply that the $\kappa$ does not exist. It means that a good dynamical
model, based on first principles, is required, in order to assess the question
whether there is or is not a pole to be associated with the $\kappa$ resonance.
This is in fact one of the aims of the present work.

Before finishing the general discussion with respect to the $\kappa$, let us
remind that this pole, together with the $\sigma(500)$, $a_0(980)$ and a
strong contribution to the $f_0(980)$, gives rise to the lightest scalar nonet
\cite{oo:99}. In this work, the whole $J^{PC}=0^{++}$ nonet was found to be
of dynamical origin, that is, due to the strong rescattering in this channel
between the pseudoscalars from the lowest order $\chi$PT amplitudes. In fact,
taking the results of ref. \cite{oo:99}, it is possible to go to the SU(3)
limit by considering the limit of equal masses for the pseudoscalars. In that
way, one can see how the poles move continuously giving rise to an octet of
degenerate scalar resonances plus a singlet.

Now we turn to the $K^*_0(1430)$ resonance. This state is clearly seen as a
steep rise in the phase, figs. 3 and 6, and as a peak structure in the modulus
of the amplitude, figs. 2 and 5. The threshold of the $K\eta'$ is very close
to the expected mass of this resonance and hence we should investigate the pole
positions both in the second $(-,+)$ and in the third $(-,-)$ Riemann sheet,
which are the nearby ones in this case. We find the following pole positions
for the fit \eqn{eq:6.11}:
\begin{equation}
\label{eq:7.2}
\begin{tabular}{cc}
Sheet & Pole Position \\
\hline
$(-,+)$ & $\sqrt{s}_{K^*_0(1430)} \,=\, (1450-i\,142)\;\mev$ \\
$(-,-)$ & $\sqrt{s}_{K^*_0(1430)} \,=\, (1358-i\,243)\;\mev$ \\
\end{tabular}
\end{equation}
There is a clear dependence in the position of the pole with respect to the
sheet. This is expected since the resonance, although originating from a tree
level pole included around 1.4 GeV, has strong unitarity corrections and is
also influenced by the nearby $K\eta'$ threshold. In any case, the pole
position found in the second Riemann sheet is very close to the effective
values of the mass $1429\pm 4\pm 5\;\mev$ and width $287\pm 10\pm 21\;\mev$
given in refs. \cite{ast:88}, deduced from a simple Breit-Wigner fit to the
experimental data. It is worth to stress that this is the lightest preexisting
$I=1/2$ scalar state.

Finally, some structure in the phase and modulus of the amplitude around
$1.7-1.9\;\gev$ is also observed in the experimental data by Aston et al.
\cite{ast:88} (solutions A and C of ref. \cite{ast:88} in figs.~5 and 6,
reminding that solutions B and D of this reference are unphysical). In ref.
\cite{ast:88}, an additional simple Breit-Wigner plus background fit was
performed for this energy region only, establishing the presence of the
$K^*_0(1950)$ resonance. The effective mass and width given in that work
varied as a function of the chosen solution for the partial wave amplitude.
For solution A, a mass of $1934\pm 8\pm 20\;\mev$ and a width of $174\pm 19 \pm
79\;\mev$ was found, whereas for the unphysical solution B the mass was
$1955 \pm 10 \pm 8\;\mev$ and the width $228 \pm 34 \pm 22\;\mev$. As explained
above, our fits were performed including data from ref.  \cite{ast:88} up to
$1.85\;\gev$. In this energy interval the solution of ref. \cite{ast:88} is
unambiguous and in fact our fits also give rise to a pole in the energy
region around $1.9\;\gev$. For the fit \eqn{eq:6.11}, the poles in the nearby
third $(-,-)$ and in the second Riemann sheet $(-,+)$ are:
\begin{equation}
\label{eq:7.3}
\begin{tabular}{cc}
Sheet & Pole Position \\
\hline
$(-,-)$ & $\sqrt{s}_{K^*_0(1950)} \,=\, (1731-i\,147)\;\mev$ \\
$(-,+)$ & $\sqrt{s}_{K^*_0(1950)} \,=\, (1908-\;\,i\, 27)\;\mev$ \\
\end{tabular}
\end{equation}
One can see a large variation in the position of this pole from one sheet to
another indicating a strong departure from the simple Breit-Wigner resonance
behaviour. Furthermore, we have checked that this resonance is an interplay
between the underlying dynamics discussed in the previous sections and
rescattering effects.\footnote{The bare parameters of the second resonance 
change considerably for the different fits and in fact for the fit
\eqn{eq:6.11} the bare mass is as high as $2.57\;\gev$.} On the other hand,
while the mass from the pole position in the second sheet is closer to the
effective one of ref. \cite{ast:88}, the width from the pole in the third
sheet is closer to the one given in that work.

Let us now come to a comparison of the pole positions resulting from different
fits of sections~4 and 6. In table~1, we give the poles corresponding to the
previously discussed resonances in the second sheet for the fits:
\eqn{eq:6.11}, \eqn{eq:6.14}, \eqn{eq:6.8}, \eqn{eq:4.4}, \eqn{eq:4.9} and
\eqn{eq:4.10}. Since the last three fits correspond to the elastic case of
section 4, only the resonances $\kappa$ and $K^*_0(1430)$ are considered for
them. In this table, we also show the residua of the poles associated with
the different channels. They are defined by the limit:
\begin{equation}
\label{eq:7.4}
r_i \;\equiv\; \lim_{s\rightarrow s_R}\sqrt{|(s-s_R) T_{ii}|}
\end{equation}
with $i$ referring to the considered channel ($i=1,2,3 \leftrightarrow
K\pi,K\eta,K\eta'$) and $s_R$ is the location of the pole for the resonance
in question. The residua presented in table~1 show that the $\kappa$ couples
strongly to the $K\pi$ state and much less to the $K\eta$ and $K\eta'$
channels. The $K^*_0(1430)$ and $K^*_0(1950)$ couple strongly with the
$K\pi$ and $K\eta'$ states with comparable strengths in both cases.  On the
other hand, the coupling of these resonances with the $K\eta$ state is much
weaker.
\begin{table}
\begin{center}
\begin{tabular}{|c|c|c|c|}
\hline
Sheet:$(-,+)$ & $\kappa$ & $K^*_0(1430)$ & $K^*_0(1950)$ \\
\hline
\eqn{eq:6.11} & $(708-i\,305)\;\mev$ & $(1450-i\,142)\;\mev$ &
 $(1908-i\,27)\;\mev$ \\
$r_{1,2,3}$ & $4.51, \; 2.36, \; 1.68$ & $4.86, \; 1.07, \; 4.12$ & 
 $5.01, \; 0.73, \; 5.97$ \\
\hline
\eqn{eq:6.14} & $(694-i\,329)\;\mev$ & $(1447-i\,160)\;\mev$ &
 $(1966-i\,50)\;\mev$ \\
$r_{1,2,3}$ & $4.83, \; 2.78, \; 1.72$ & $5.32, \; 1.29, \; 4.71$ &
 $5.75, \; 0.99, \; 7.03$ \\
\hline
\eqn{eq:6.8} & $(714-i\,301)\;\mev$ & $(1429-i\,146)\;\mev$ &
 $(1880-i\,112)\;\mev$ \\
$r_{1,2,3}$ & $4.31, \; 1.55, \; 1.19$ &  $4.82, \; 0.85,\; 2.67$ & 
 $4.76, \; 0.91, \; 4.20$ \\
\hline
\eqn{eq:4.4} & $(701-i\,320)\;\mev$ & $(1437-i\,70)\;\mev$ & \\
$r_1$ & 4.79 & 3.53 & Elastic Case \\
\hline
\eqn{eq:4.9} & $(709-i\,306)\;\mev$ & $(1386-i\,48)\;\mev$ & \\
$r_1$ & 4.74 & 2.85 & Elastic Case \\
\hline
\eqn{eq:4.10} & $(684-i\,279)\;\mev$ & $(1407-i\,140)\;\mev$ & \\
$r_1$ & 4.10 & 4.21 & Elastic Case \\
\hline
\end{tabular}
\caption[]{A collection of the pole positions and residua $r_i$ for the fits
of eqs. \eqn{eq:6.11}, \eqn{eq:6.14}, \eqn{eq:6.8}, \eqn{eq:4.4}, \eqn{eq:4.9}
and \eqn{eq:4.10} in the second Riemann sheet. The $r_i$ are given in units
of $M_\pi$.}
\end{center}
\end{table}

When comparing the different fits in table~1, one sees that the properties of
the $\kappa$ meson are rather stable when changing from one fit to another.
For the $K^*_0(1430)$ one observes a common trend in the inelastic fits, as
well as in the elastic one of eq.~\eqn{eq:4.10}, to give a pole around
$(1.44-i\,0.15)\;\gev$. There is, however, a large difference between the pole
position and the residua $r_1$ corresponding to the elastic fit \eqn{eq:4.9}
and the ones given by the inelastic fits. On the other hand, the values from
the fit \eqn{eq:4.4} lie in an intermediate region. We will further discuss
these points after introducing table~2 below.

For the $K^*_0(1950)$ the situation is more unstable when comparing different
fits, particularly with respect to the imaginary part of the pole position in
the second sheet. However, as can be seen in figs. 5 and 6, all the inelastic
fits result in rather similar curves. This clearly implies that the role of
the $K^*_0(1950)$ is hidden by a very large background and hence one needs a
very precise approach for such large energies around $2\;\gev$ in order to
extract the resonance-pole parameters. This is clearly out of the scope of the
present paper because for such high energies, additional multiparticle states
can play a significant role and also the contributions of the unphysical cuts
become increasingly large. Let us note that for the fit \eqn{eq:6.8} the pole
position corresponding to this resonance is very similar to the mass and
width obtained in ref. \cite{ast:88}, the only reference considered in the
Review of Particle Physics \cite{pdg:98} for the $K^*_0(1950)$.

\begin{table}
\begin{center}
\begin{tabular}{|c|c|c|}
\hline
Sheet:$(-,-)$ & $K^*_0(1430)$ & $K^*_0(1950)$ \\
\hline
\eqn{eq:6.11} & $(1358-i\,243)\;\mev$ & $(1731-i\,147)\;\mev$ \\
$r_{1,2,3}$ & $5.56, \; 1.65, \; 4.41$ & $3.67, \; 0.24, \; 4.84$ \\
\hline
\eqn{eq:6.14} & $(1299-i\,255)\;\mev$ & $(1700-i\,151)\;\mev$ \\
$r_{1,2,3}$ & $5.52, \; 1.44, \; 4.07$ & $3.79, \; 0.24, \; 4.62$ \\
\hline
\eqn{eq:6.8} & $(1394-i\,188)\;\mev$ & $(1772-i\,172)\;\mev$ \\
$r_{1,2,3}$ & $5.18, \; 0.84, \; 3.19$ & $4.24, \; 0.58, \; 4.42$ \\
\hline
\end{tabular}
\caption[]{A collection of the pole positions and residua $r_i$ for the
inelastic fits of eqs. \eqn{eq:6.11}, \eqn{eq:6.14} and \eqn{eq:6.8} in the
third Riemann sheet. The $r_i$ are given in units of $M_\pi$.}
\end{center} 
\end{table}

Table~2 is analogous to table~1, but now considering the third Riemann sheet
$(-,-)$. Therefore, in this second table, we do not show the $\kappa$
resonance and we only consider inelastic fits. The third sheet is in principle
the closest one to the physical $s$-axis when one is well above the $K\eta'$
threshold. This is clearly the situation for the $K^*_0(1950)$. For the
$K^*_0(1430)$ one also needs to consider the second sheet $(-,+)$ discussed
above, since we are very close to the $K\eta'$ threshold. From table~2, one
infers that for this latter case the pole associated with the $K^*_0(1430)$
in the third sheet is very broad, but with similar residua to the one in the
second sheet, and thus one would expect that the pole driving the effects of
the $K^*_0(1430)$ on the physical axis should correspond to the one in the
second sheet. In fact, the masses and widths of the $K^*_0(1430)$ derived from
the second sheet poles listed in table~1, corresponding to the inelastic fits
and also to the elastic one \eqn{eq:4.10}, are in fairly good agreement with
the effective values of ref. \cite{ast:88}.

Concerning the $K^*_0(1950)$ resonance poles, there is no clear and simple 
criterion in order to decide which pole, the one in the second or in the third
sheet, is the leading one. In fact, when comparing with the experimental data
of solutions A and C of ref. \cite{ast:88}, fig.~5, one observes a dip in the
modulus of the amplitude around 1.7 GeV. This precisely corresponds to the
real part of the pole position corresponding to the $K^*_0(1950)$ in the third
Riemann sheet, table~2. One the other hand, when looking at the phase shifts
for the same solutions in fig.~6, one sees a local maximum around $1.9\;\gev$,
which coincides with the real part of the poles in the second sheet, table~1.
Furthermore, as discussed above, in the region of the $K^*_0(1950)$ there is
a large background that could mimic the effects of the broad and lighter pole
found in the third sheet.

\newsection{Conclusions}

S-wave scattering of the pseudoscalar Goldstone bosons is a good testing
ground for our understanding of strong interactions in the non-perturbative
region. At low energies, the Goldstone dynamics is constrained by the chiral
symmetry properties of QCD. The $\chi$PT Lagrangian contains all relevant
information on the infrared singularities (thresholds, cuts, etc.), which
allows to reconstruct the scattering S-matrix elements up to local subtraction
polynomials. The structure of these local terms is also governed by chiral
symmetry, but their coefficients, the chiral couplings, encode the
short-distance information.

The $\chi$PT predictions can be extrapolated to higher energies by taking into
account the leading massive singularities, i.e. incorporating the lightest
resonance poles through a chiral-symmetric Effective Field Theory with
resonance fields as explicit degrees of freedom \cite{egpr:89,eglpr:89}.
Combined with large-$N_c$ arguments to organise a well-defined perturbative
approach, this procedure provides a good understanding of many low and
intermediate energy phenomena. However, it is not enough to achieve a correct
description of the scalar sector, because the S-wave rescattering of two
pseudoscalars is very strong, making necessary to perform a resummation of
chiral-loop corrections in order to satisfy unitarity \cite{gp:97,gue:98,
oop:99,oo:99,oor:00}.

In this paper, we have tried to describe the S-wave $K\pi$ scattering up to
$2\;\gev$, through a unitarisation of the resonance chiral Lagrangian
predictions. In the elastic region a reasonably good description can be
achieved for the dominant $I=1/2$ amplitude, which satisfies all low-energy
and high-energy constraints. However, the more exotic $I=3/2$ sector remains
problematic, in spite of the clear improvement provided by the unitarisation
procedure. Being pure background (no resonances), and with very large
contributions from crossed channel dynamics, it is more difficult to get
a good high-energy extrapolation for the $I=3/2$ amplitude since many missing
small corrections could be numerically important. Nevertheless, the present
discrepancies among different sets of data prevent us from reaching any firm
conclusion; in fact, the older $I=3/2$ data appears to be better described
within the chiral framework.

Above roughly 1.3 GeV, the $I=1/2$ amplitude contains inelastic contributions.
We have incorporated the leading two-body $K\eta$ and $K\eta'$ modes, through
a coupled channel analysis. The necessary input are the five independent 
$T_{K\pi\to K\eta_i}$ and $T_{K\eta_i\to K\eta_j}$ amplitudes, which we have
computed at the required order.

The inelastic fit faces the problem of a very unsatisfactory experimental
status. The data only determines the absolute value of the scattering
amplitude, leaving $2^L$ discrete ambiguous solutions to the partial wave
decomposition with $L$ being the maximal angular momentum contributing at a
given energy. Moreover, different experiments give rise to different solutions
making rather unclear the comparison with theory. Even the masses and widths
of the dominant scalar resonances are not clearly established.

We have combined the known theoretical ingredients in order to see which
experimental solutions are preferred. Up to 1.85 GeV, the Aston data
\cite{ast:88,awa:86} corresponds to solution B of Estabrooks \cite{est:78}.
Arguing that this should be the physical solution, we have performed several
fits in the respective region and studied their behaviour at higher energies
where the data ambiguities are more severe. Although some of the solutions of
ref. \cite{ast:88} for energies higher than $1.85\;\gev$ are shown to be
unphysical, since they violate unitarity and can be discarded, the large
experimental errors do not allow to unambiguously select the physical one.

A similar fit quality could be achieved with rather different values of the
input parameters. Nevertheless, when imposing the high-energy constraints
discussed above, definite values for some of the parameters can be given.
In addition, assuming saturation of the low-energy constants $L_5$ and $L_8$
by scalar-resonance contributions, we were able to estimate values for these
parameters which are compatible with the very recent work \cite{abt:99,abt:00}.
We have also used our fits to study the position of the T-matrix poles in the
complex plane, aiming to determine the mass and width of the dominant scalar
resonances.

Clearly, better experimental data are needed. Our results provide a convenient
theoretical framework to be used in future experiments for analysing the data
and resolving possible ambiguities. Even with all present shortcomings,
our analysis considerably improves the knowledge of the S-wave $K\pi$
scattering amplitude. In a forthcoming publication \cite{jop:00b}, we will
use this information to perform a detailed investigation of the $K\pi$ scalar
form factor up to 2 GeV. We expect that this could help to improve the
determination of the strange quark mass from QCD sum rules for the
strangeness-changing scalar current \cite{nprt:83,mm:93,cdps:95,cfnp:97,
jam:97,bgm:98}.

\vspace{6mm} \noindent
{\Large\bf Acknowledgements}

\vspace{3mm} \noindent
We would like to thank B.~Ananthanarayan and P.~B\"uttiker for
calling our attention to ref.~\cite{ls:78} as well as H.~G.~Dosch and
U.-G.~Mei\ss ner for critically reading the manuscript.
This work has been supported in part by the German-Spanish Cooperation
Agreement HA97-0061, by the European Union TMR Network EURODAPHNE
(ERBFMX-CT98-0169), and by DGESIC (Spain) under the grants no. PB97-1261
and PB96-0753. M. J. would like to thank the Deutsche Forschungsgemeinschaft
for support.

\newpage

\appendix{\LARGE\bf Appendices}

\newsection{Loop functions}

For completeness, we tabulate here the one-loop functions \cite{gl:85}
appearing in the unitarity chiral correction of eq.~\eqn{eq:2.5}. The basic
integral is given by:
\begin{eqnarray}
\bar{J}_{PQ}(s) &\equiv & -{1\over 16\pi^2}\,\int_0^1 dx \,
\log{\left[ {M_P^2 - s x (1-x) - \Delta_{PQ} x \over 
   M_P^2  - \Delta_{PQ} x} \right]}
\\ & = &
{1\over 32\pi^2}\, \left\{ 2 + 
  \left( {\Delta_{PQ} \over s} - {\Sigma_{PQ}\over \Delta_{PQ}}\right)
  \log{\left({M_Q^2\over M_P^2}\right)} - {\lambda_{PQ}\over s}
  \log{\left[ {\left( s+\lambda_{PQ}\right)^2 - \Delta_{PQ}^2\over
     \left( s-\lambda_{PQ}\right)^2 - \Delta_{PQ}^2}\right]} \right\} \,, \nn
\end{eqnarray}
with
$$
\Sigma_{PQ} \equiv M_P^2 + M_Q^2  \; ; \quad
\Delta_{PQ} \equiv M_P^2 - M_Q^2  \; ; \quad
\lambda_{PQ}^2\equiv  \left[ s - \left( M_P+M_Q\right)^2\right]
  \left[ s - \left( M_P-M_Q\right)^2\right] \; .
$$

From $\bar{J}_{PQ}(s)$, one defines the related quantities:
\begin{equation}
J^r_{PQ}(s) \equiv \bar{J}_{PQ}(s) - 2 k_{PQ}(\mu) \; ; \quad
K_{PQ}(s) \equiv  {\Delta_{PQ} \over 2 s}\,  \bar{J}_{PQ}(s) \; ; \quad
L_{PQ}(s) \equiv  {\Delta_{PQ}^2 \over 4 s}\,  \bar{J}_{PQ}(s) \; ;
\end{equation}
and
\begin{equation}
M^r_{PQ}(s) \equiv {1\over 12 s} \left( s - 2 \Sigma_{PQ}\right)
  \bar{J}_{PQ}(s) 
  + {\Delta_{PQ}^2 \over 3 s^2}\, \left[ \bar{J}_{PQ}(s) - s J'_{PQ}\right]
  - {1\over 6} k_{PQ}(\mu) + {1\over 288 \pi^2} \, .
\end{equation}
Here,
\begin{eqnarray}
k_{PQ}(\mu) & \equiv & {1\over 32\pi^2\Delta_{PQ}}\, \Big[
   M_P^2 \log{\left(M_P^2/\mu^2\right)} - M_Q^2 \log{\left(M_Q^2/\mu^2\right)}
   \Big] \,  ,
\nn\\
\smvs
J'_{PQ} & \equiv & {1\over 32\pi^2}\, \left\{
  {\Sigma_{PQ}\over \Delta_{PQ}^2 } + 2 {M_P^2 M_Q^2 \over \Delta_{PQ}^3 }\,
  \log{\left( { M_Q^2\over M_P^2}\right)} \right\} \, ,
\end{eqnarray}
and $\mu$ denotes the renormalisation scale.

\newpage

\newsection{\boldmath $\eta$-$\eta'$ mixing\unboldmath}

Let us consider the two-dimensional space of isoscalar pseudoscalar
mesons. We collect the SU(3) octet and singlet fields in the doublet
$\eta_B^T \equiv (\eta_8, \eta_1)$. The quadratic term in the Lagrangian
takes the form
\begin{equation}
\cL \; = \; {1\over 2}\,\partial_\mu \eta_B^T\, \cK \,\partial^\mu \eta_B -
{1\over 2}\,\eta_B^T\, \cM^2 \, \eta_B \, ,
\end{equation}
with
\begin{equation}
\cK \; = \; \left( 
\begin{array}{cc}
1 + \epsilon_8 & \delta \\
        \delta & 1 + \epsilon_1
\end{array}
\right) \,, \qquad
\cM^2 \; = \; \left( 
\begin{array}{cc}
   M_8^2 &  M_{18}^2 \\
M_{18}^2 &  M_1^2
\end{array}
\right) \,.
\end{equation}

The coupling to the scalar resonances generates a $\cK\not= I_2$ kinetic
matrix,
\begin{equation}
\epsilon_8 \; = \; {8 c_d c_m\over f^2 M_S^2} \, \dotM{8} \,, \qquad
\epsilon_1 \; = \; {8 c_d c_m\over f^2 M_S^2} \, \dotM{1} \,, \qquad
\delta \; = \; -\,{2\sqrt{2}\over 3}\,{8 c_d c_m\over f^2 M_S^2}\,
\Delta_{K\pi} \,,
\end{equation}
and modifies the lowest-order values of the mass-matrix elements:
\begin{eqnarray}
M_8^2 &=& \dotM{8} + {8 c_m^2\over f^2 M_S^2}\biggl( M_\pi^4 +
\frac{8}{3}\, M_K^2 \Delta_{K\pi} \biggr) \,, \nn \\
\smvs
M_1^2 &=& M_0^2 + \dotM{1} + {8 c_m^2\over f^2 M_S^2} \biggl( M_\pi^4 +
\frac{4}{3}\, M_K^2 \Delta_{K\pi} \biggr) \,, \\
\smvs
M_{18}^2 &=& -\,{2\sqrt{2}\over 3}\,\stackrel{0}{\Delta}_{K\pi} \biggl( 1 +
{8 c_m^2\over f^2 M_S^2}\, 2 M_K^2 \biggr) \,. \nn
\end{eqnarray}
Here, $M_0^2$ denotes the ${\rm U}(1)_A$ anomaly contribution to the
$\eta_1$ mass, $\dotM{i}$ ($i=1,8$) the $\cO(p^2)$ contributions to the
singlet and octet isoscalar masses,
\begin{equation}
\dotM{8} \; = \; {1\over 3} \biggl( 4 \dotM{K} - \dotM{\pi} \biggr) \,,
\qquad
\dotM{1} \; = \; {1\over 3} \biggl( 2 \dotM{K} + \dotM{\pi} \biggr) \,,
\end{equation}
and $\stackrel{0}{\Delta}_{K\pi} \equiv \dotM{K} - \dotM{\pi}$, with
$\dotM{\pi}$ and $\dotM{K}$ the pion and kaon masses at $\cO(p^2)$ in the
chiral expansion.

To first order in $\epsilon_1$, $\epsilon_8$ and $\delta$, the kinetic matrix
$\cK$ can be diagonalised through the following field redefinition:
\begin{equation}
\eta_B \; = \; Z^{1/2} \cdot \hat{\eta} \equiv Z^{1/2} \cdot
\left( \begin{array}{c} \hat{\eta}_8 \\ \hat{\eta}_1 \end{array} \right) \,,
\quad
{Z^{1/2}}^T \cdot \cK \cdot Z^{1/2} \; = \; I_2 \,, \quad
Z^{1/2} \; = \; \left( 
\begin{array}{cc}
1 - {\epsilon_8\over 2} & -\,{\delta\over 2} \\ 
-\,{\delta\over 2} & 1 - {\epsilon_1\over 2} 
\end{array} \right) \,.
\end{equation}
In the $\hat{\eta}$ basis the mass matrix takes the form
\begin{equation}
\hat{\cM} = {Z^{1/2}}^T \cdot \cM \cdot Z^{1/2} \,,
\end{equation}
where
\begin{eqnarray}
\hat{M}_8^2 &=&  \dotM{8} + {8 c_m(c_m-c_d)\over f^2 M_S^2}\, \biggl(
M_\pi^4 + \frac{8}{3}\, M_K^2 \Delta_{K\pi} \biggr) \,, \nn \\
\hat{M}_1^2 &=& M_0^2 \biggl( 1 - {8 c_m c_d \over f^2 M_S^2}\dotM{1} \biggr)
+ \dotM{1} + {8 c_m(c_m-c_d)\over f^2 M_S^2}\, \biggl( M_\pi^4 +
\frac{4}{3}\, M_K^2 \Delta_{K\pi} \biggr) \,, \\
\hat{M}_{18}^2 &=& -\,{2\sqrt{2}\over 3}\,\stackrel{0}{\Delta}_{K\pi} \, 
\biggl( 1 + {8 c_m(c_m-c_d)\over f^2 M_S^2}\,2 M_K^2 -
{4 c_m c_d \over f^2 M_S^2}\, M_0^2 \biggr) \,.  \nn
\end{eqnarray}

The physical mass eigenstates are obtained by diagonalising the matrix
$\hat{\cM}$ with an orthogonal transformation,
\begin{equation}
\hat{\cM} \; = \; \cR^T \cdot \cM_D \cdot \cR \,, \quad
\hat{\eta} \; = \; \cR^T \cdot \eta_P \; \equiv \; \cR^T \cdot \left(
\begin{array}{c} \eta \\ \eta' \end{array} \right) \,, \quad
\cR \; \equiv \; \left( \begin{array}{cc}
\cos{\theta_P} & -\,\sin{\theta_P} \\ \sin{\theta_P} & \cos{\theta_P}
\end{array} \right) \,.
\end{equation}
One gets the relations
\begin{equation}
M_{\eta'}^2 + M_\eta^2 \; = \; \hat{M}_1^2 + \hat{M}_8^2 \,, \qquad
M_{\eta'}^2 - M_\eta^2 \; = \; \sqrt{(\hat{M}_1^2 - \hat{M}_8^2)^2 +
4 \hat{M}_{18}^4} \,,
\end{equation}
and
\begin{equation}
\tan{\theta_P} \; = \; {\hat{M}_8^2 - M_\eta^2\over \hat{M}_{18}^2} \; = \;
{M_{\eta'}^2 - \hat{M}_1^2 \over \hat{M}_{18}^2} \; = \; 
{\hat{M}_{18}^2\over M_{\eta'}^2 - \hat{M}_8^2} \; = \;
{\hat{M}_{18}^2\over \hat{M}_1^2 -  M_\eta^2} \, .
\end{equation}

\newpage

\newsection{Inelastic channels}

\begin{eqnarray}
\label{eq:C1}
T_{ K\pi\to K\eta_1} &\!\!=\!\!&  \frac{1}{\sqrt{2} f_K^2}
\Biggl\{ {2 M_K^2 + M_\pi^2\over 3} + \frac{2 c_m c_d\,\DelKP}{3 f^2 M_S^2}\,
(3 t - p_\eta^2 + M_\pi^2) - \frac{8 c_m^2\DelKP^2}{3 f^2 M_S^2}
\nn \\ \smvs
& & + \,{2\over f^2}\biggl[ {c_d\, s + (c_m-c_d)\SigKP\over
M_{K^*_0}^2 - s} \Big[ c_d\, (s - p_\eta^2 - M_K^2) + 2 c_m M_K^2 \Big] \nn \\
\smvs
&& \hspace{7mm} + \,{c_d\, u + (c_m-c_d)\SigKP\over M_{K^*_0}^2 - u}
\Big[ c_d\, (u - p_\eta^2 - M_K^2 ) + 2 c_m M_K^2 \Big] \nn \\
\smvs
&& \hspace{7mm} + \,{c_d\, t + 2 (c_m-c_d) M_K^2\over M_{a_0}^2 - t} \Big[
c_d\,(t - p_\eta^2 - M_\pi^2) + 2 c_m M_\pi^2 \Big] \biggr]\Biggr\} \,,
\end{eqnarray}

\begin{eqnarray}
\label{eq:C2}
T_{K\pi\to K\eta_8 } &\!\!=\!\!& \frac{1}{4f_K^2}\,
\Biggl\{ 3 t - p_\eta^2 - {8\over 3} M_K^2 - {1\over 3} M_\pi^2 +
\frac{4 c_m c_d\,\DelKP}{3f^2 M_S^2}\,( 3t + 2p_\eta^2 - 2 M_\pi^2 )
-\frac{4 c_m^2\DelKP^2}{3f^2 M_S^2} 
\nn \\ \smvs
&& +\,{2\over f^2}\biggl[ {c_d\, s + (c_m-c_d)\SigKP\over
M_{K^*_0}^2 - s} \Big[ c_m\,(3 M_\pi^2 - 5 M_K^2) -
c_d\, (s - p_\eta^2 - M_K^2)\Big] \nn \\
\smvs
&& \hspace{7mm} + \,{c_d\, u + (c_m-c_d)\SigKP\over M_{K^*_0}^2 - u}
\Big[ c_m\, (3 M_\pi^2 - 5 M_K^2) - c_d\, (u - p_\eta^2 - M_K^2)\Big] \nn \\
\smvs
&& \hspace{4mm} + \,2\, {c_d\, t + 2 (c_m-c_d) M_K^2 \over M_{a_0}^2 - t}
\Big[ 2 c_m M_\pi^2 + c_d\, (t - p_\eta^2 - M_\pi^2) \Big] \biggr] \nn \\
\smvs
&& + \,{3 G_V^2\over f^2}\biggl[ { s (t-u) - (p_\eta^2-M_K^2)\DelKP\over
M_{K^*}^2 - s} + { u (t-s) - (p_\eta^2 - M_K^2)\DelKP \over M_{K^*}^2 - u}
\biggr] \Biggr\} \,,
\end{eqnarray}

\begin{eqnarray}
\label{eq:C3}
T_{K\eta_1\to K\eta_1} &\!\!=\!\!& {4\over 3 f_K^2} \,\Biggl\{
\,{1\over 2}\,M_K^2 - {4 c_m c_d\,\DelKP^2\over 3 f^2 M_S^2} \nn \\
\smvs
&& \hspace{-3mm}+\,{1\over f^2}\biggl[\,{1 \over M_{K^*_0}^2 - s} \,
\Big[ c_d\, (s - k_\eta^2 - M_K^2) + 2 c_m M_K^2 \Big]
\Big[ c_d\, (s - {k'_\eta}^2 - M_K^2) + 2 c_m M_K^2 \Big] \nn \\
\smvs
&& \hspace{4.5mm} +\,{1 \over M_{K^*_0}^2 - u} \,
\Big[ c_d\, (u - k_\eta^2 - M_K^2) + 2 c_m M_K^2 \Big]
\Big[ c_d\, (u - {k'_\eta}^2 - M_K^2) + 2 c_m M_K^2 \Big] \nn \\
\smvs
&& \hspace{4.5mm} +\,{1 \over M_{S_1}^2 - t} \,
\Big[ c_d\, t + 2 (c_m -c_d) M_K^2 \Big]
\Big[ c_d\, (t-k_\eta^2-{k'_\eta}^2) + \msmall{2\over 3}\,
c_m (2 M_K^2+M_\pi^2)\Big] \nn \\ \smvs
&& \hspace{1mm} +\, {2\over 3}\, {c_m\,\DelKP \over M_{S}^2 - t} \,
\Big[ c_d\, t + 2 (c_m - c_d) M_K^2 \Big]\,\biggr]\,\Biggr\} \,,
\end{eqnarray}

\begin{eqnarray}
\label{eq:C4}
T_{K\eta_1\to K\eta_8} &\!\!=\!\!& -\,{\sqrt{2}\over 3 f_K^2}\,
\Biggl\{\,M_K^2 - \msmall{1\over 2} M_\pi^2 
- {c_m c_d\,\DelKP\over 3 f^2 M_S^2}
(9 t - 3 k_\eta^2 + 3 {k'_\eta}^2 + 16 \DelKP )
+ {4 c_m^2\DelKP^2\over f^2 M_S^2} \nn \\
\smvs
&& \hspace{-18mm}+\,{1\over f^2}\biggl[\, {1 \over M_{K^*_0}^2 - s} \,
\Big[ c_d\, (s - k_\eta^2 - M_K^2) + 2 c_m M_K^2 \Big]
\Big[ c_d\, (s - {k'_\eta}^2 - M_K^2) + c_m ( 5 M_K^2 - 3 M_\pi^2) \Big] \nn \\
\smvs
&& \hspace{-10.5mm}+\,{1 \over M_{K^*_0}^2 - u} \,
\Big[ c_d\, (u - k_\eta^2 - M_K^2) + 2 c_m M_K^2 \Big]
\Big[ c_d\, (u - {k'_\eta}^2 - M_K^2) + c_m ( 5 M_K^2 - 3 M_\pi^2)\Big] \nn \\
\smvs
&& \hspace{-10.5mm}+\,{1 \over M_{S}^2 - t} \,
\Big[ c_d\, t + 2 (c_m -c_d) M_K^2 \Big]
\Big[ c_d\, (t - k_\eta^2 - {k'_\eta}^2) + \msmall{2\over 3}\,
c_m (4 M_K^2 - M_\pi^2)\Big] \nn \\
\smvs
&& \hspace{-16mm}+\,{16\over 3}\, {c_m\,\DelKP \over M_{S_1}^2 - t} \,
\Big[ c_d\, t + 2 (c_m - c_d) M_K^2 \Big]\,\biggr]\,\Biggr\} \,,
\end{eqnarray}

\begin{eqnarray}
\label{eq:C5}
T_{K\eta_8\to K\eta_8} &\!\!=\!\!& {1\over 6 f_K^2}\,\Biggl\{\,
{1\over 2}\, (9 t - 3 k_\eta^2 - 3 {k'_\eta}^2 - 2 M_\pi^2) 
- {2 c_m c_d\,\DelKP\over 3 f^2 M_S^2} (9 t + 32 \DelKP )
+ {14 c_m^2\DelKP^2\over f^2 M_S^2} \nn \\
\smvs
&& \hspace{-28mm}+\,{1\over f^2}\biggl[\,{1 \over M_{K^*_0}^2 - s} \,
\Big[ c_d\, (s - k_\eta^2 - M_K^2) + c_m ( 5 M_K^2 - 3 M_\pi^2) \Big]
\Big[ c_d\, (s - {k'_\eta}^2 - M_K^2) + c_m ( 5 M_K^2 - 3 M_\pi^2) \Big] \nn \\
\smvs
&& \hspace{-20.5mm}+\,{1 \over M_{K^*_0}^2 - u} \,
\Big[ c_d\, (u - k_\eta^2 - M_K^2) + c_m ( 5 M_K^2 - 3 M_\pi^2)\Big]
\Big[ c_d\, (u - {k'_\eta}^2 - M_K^2) + c_m ( 5 M_K^2 - 3 M_\pi^2)\Big] \nn \\
\smvs
&& \hspace{-8mm}+\,{2 \over M_{S}^2 - t} \,
\Big[ c_d\, t + 2 (c_m -c_d) M_K^2 \Big]
\Big[ c_d\, (t - k_\eta^2 - {k'_\eta}^2) + \msmall{2\over 3}\,
c_m (8 M_K^2 - 5 M_\pi^2)\Big] \nn \\
\smvs
&& \hspace{-8mm}+\,{8\over M_{S_1}^2 - t} \,
\Big[ c_d\, t + 2 (c_m - c_d) M_K^2 \Big]
\Big[ c_d\, (t - k_\eta^2 - {k'_\eta}^2) + \msmall{2\over 3}\,
c_m (4 M_K^2 - M_\pi^2)\Big]\,\biggr] \nn \\
\smvs
&& \hspace{-20.5mm}+\,{9\,G_V^2 \over 2 f^2} \, \Biggl[
{s (t-u) + (k_\eta^2 - M_K^2) ({k'_\eta}^2 - M_K^2) \over  M_{K^*}^2 - s} +
{u (t-s) + (k_\eta^2 - M_K^2) ({k'_\eta}^2 - M_K^2) \over  M_{K^*}^2 - u}
\Biggr]\,\Biggr\} \, .
\end{eqnarray}

Several remarks concerning the above expressions are in order.
$p_\eta$, $k_\eta$ and $k'_\eta$ are momenta of the $\eta_i$'s.
In eqs.~\eqn{eq:C3} to \eqn{eq:C5}, $k_\eta$ refers to the initial state
$\eta_i$ and $k'_\eta$ to the final state $\eta_j$. For example in the
calculation of the amplitude $T_{K\eta\to K\eta'}$, we have ${k_\eta}^2=
M_{\eta}^2$ and ${k'_\eta}^2=M_{\eta'}^2$.

\newsection{Input parameters}

For the convenience of the reader, in this appendix we collect the values of
all constant input parameters used in our numerical analysis:
\begin{displaymath}
M_\pi  \; = \; 138.0\;\mev \,, \qquad
M_K    \; = \; 495.7\;\mev \,,
\end{displaymath}
\begin{displaymath}
M_\eta \; = \; 547.3\;\mev \,, \qquad
M_{\eta'} \; = \; 957.8\;\mev \,,
\end{displaymath}
\begin{displaymath}
f_\pi  \; = \;  92.4\;\mev \,, \qquad
f_K    \; = \; 112.7\;\mev \,,
\end{displaymath}
\begin{displaymath}
M_\rho  \; = \; 770.0\;\mev \,, \qquad
M_{K^*} \; = \; 896.1\;\mev \,,
\end{displaymath}
\begin{displaymath}
M_{V'}  \; = \; 1440\;\mev \,, \qquad
M_{V''} \; = \; 1710\;\mev \,,
\end{displaymath}
\begin{displaymath}
G_V \; = \; f_\pi/\sqrt{2} \; = \; 65.3\;\mev \,,
\end{displaymath}
\begin{displaymath}
G_{V'}  \; = \; 12\;\mev \,, \qquad
G_{V''} \; = \; 23\;\mev \,,
\end{displaymath}
\begin{displaymath}
M_S \; = \; M_{S_1} \,, \qquad
M_{S'} \; = \; M_{S_1'} \,,
\end{displaymath}
\begin{displaymath}
\sin\theta \; = \; -\,1/3 \;\approx\; 20^\circ \,.
\end{displaymath}

\newpage

\end{document}